\RequirePackage{fix-cm}

\documentclass[twocolumn,epjc3]{svjour3}

\smartqed  

\RequirePackage{graphicx}
\RequirePackage{amsmath}
\RequirePackage{amssymb}
\RequirePackage{bm}
\RequirePackage{physics}
\RequirePackage{bbm}
\RequirePackage{braket}
\RequirePackage{slashed}
\RequirePackage{amsfonts}
\RequirePackage{float}
\RequirePackage{tikz}
\usetikzlibrary{positioning}
\RequirePackage[compat=1.1.0]{tikz-feynman}

\RequirePackage[hyphens]{url} 
\RequirePackage{hyperref}
\hypersetup{breaklinks=true, colorlinks=true, linkcolor=red, filecolor=black, citecolor=blue}
\RequirePackage{xurl} 
\RequirePackage[english]{babel}
\usepackage[final]{microtype}
\begin{document}

\title{Dynamical mass generation and critical behavior in pseudo-Proca quantum electrodynamics}

\author{Helio G. Barroso\thanksref{e1,addr1}
        \and
        Van S\'ergio Alves\thanksref{e2,addr1}
        \and
        Leandro O. Nascimento\thanksref{e3,addr1,addr2}}

\thankstext{e1}{e-mail: helio.barroso@icen.ufpa.br}
\thankstext{e2}{e-mail: vansergi@ufpa.br}
\thankstext{e3}{e-mail: lon@ufpa.br}

\institute{Faculdade de F\'isica, Universidade Federal do Par\'a, Avenida Augusto Correa 01, 66075-110, Bel\'em, Par\'a, Brazil \label{addr1}
           \and
           Universidade Federal de Campina Grande, Rua Apr\'igio Veloso 882, 58429-900, Campina Grande, Para\'iba, Brazil \label{addr2}
}

\date{Received: date / Accepted: date}

\maketitle

\begin{abstract}
We investigate dynamical mass generation in pseudo-Proca quantum electrodynamics (PPQED) by means of Schwinger--Dyson equations in rainbow-quenched and unquenched truncations. The pseudo-Proca screening scale $m$, together with the fine-structure constant $\alpha$, the flavor number $N$, and the ultraviolet cutoff $\Lambda$, control the critical thresholds for chiral symmetry breaking in the reduced $(2+1)$D theory. Within these truncation schemes, we obtain analytical estimates for both the critical coupling $\alpha_c(m,\Lambda)$ and the critical number of fermion flavors $N_c(m,\Lambda,g)$, and show that increasing $m$ suppresses dynamical mass generation through Yukawa screening. We further assess the robustness of this physical picture by incorporating a Ball--Chiu vertex construction, which modifies the quantitative values of the critical parameters while preserving their qualitative dependence on the screening scale $m$. In addition, within the static approximation, the anisotropic extension indicates that the critical coupling decreases as the ratio $v_F/c$ increases. Taken together, our results suggest that the scale $m$ plays an important role in modulating criticality in PPQED. Furthermore, the Ball-Chiu vertex construction shows that vertex corrections modify the quantitative values of the critical parameters while preserving the screening-driven suppression of dynamical mass generation in $(2+1)$D.

\end{abstract}

\section{Introduction}
\label{sec:introduction}

Quantum electrodynamics (QED) is one of the cornerstones of modern physics, providing a remarkably accurate description of electromagnetic interactions between charged fermions and gauge fields. Its modern formulation emerged from the seminal works of Dirac, Tomonaga, Schwinger, Feynman, and Dyson~\cite{Dirac:1927dy,Dirac:1928hu,Tomonaga:1946zz,Schwinger1948,Feynman1949,Dyson1949}. Although the standard QED is formulated in $(3+1)$D, lower-dimensional realizations are of continuing interest because they provide effective descriptions of planar systems in which fermionic degrees of freedom are confined to a two-dimensional spatial manifold.

A particularly important example is pseudo-quantum electrodynamics (PQED), a $(2+1)$D effective field theory that describes planar fermions interacting through the full electromagnetic field while preserving gauge invariance and the long-range Coulomb interaction characteristic of $(3+1)$D electrodynamics~\cite{marino1993quantum,Gorbar2001,Kovner1990}. In its non-perturbative regime, PQED has been successfully analyzed through Schwinger–Dyson equations (SDEs), which reveal critical thresholds for dynamical mass generation and chiral symmetry breaking~\cite{alves2013chiral}. The framework has also been extended to include four-fermion interactions~\cite{junior2017}, finite-temperature effects~\cite{fernandez2021dynamical}, and excitonic phenomena in transition-metal dichalcogenides, with predictions in good agreement with experiment~\cite{Marino_2018_2dmaterials}.

Massive gauge sectors, on the other hand, play a central role in several areas of high-energy and condensed-matter physics. In relativistic field theory, gauge-boson masses arise either dynamically or through explicit mass terms, with well-known realizations in Proca- and Stueckelberg-type constructions~\cite{higgs1964broken,weinberg1967model,gunion2018higgs,stuckelberg1938interaction,Ruegg2004}. When the corresponding gauge dynamics is dimensionally reduced to a planar setting, one obtains pseudo-Proca quantum electrodynamics (PPQED), also referred to as non-local Proca QED~\cite{alves2018two,ozela2023effective}. This theory preserves gauge invariance while introducing a non-local kernel that interpolates between Coulomb-like and Yukawa-screened interactions. Previous studies of PPQED have mainly focused on perturbative aspects of the model; by contrast, its non-perturbative chiral properties remain largely unexplored.

This problem is physically relevant because the presence of a screening scale $m$ modifies the infrared structure of the interaction and may therefore alter the onset of chiral symmetry breaking in a non-trivial way. In low-energy settings, Yukawa-type interactions are also relevant for modeling screened electromagnetic interactions in two-dimensional materials and for characterizing the response of so-called Proca metamaterials~\cite{Said2021}. In the former case, many-body effects may induce screening of the Coulomb interaction. In the latter, possible classical analogues of Proca electrodynamics may offer qualitative motivation for interpreting the role of an effective mass scale in Yukawa-like interactions.

A central non-perturbative phenomenon in QED and related gauge theories is dynamical mass generation, whereby fermions acquire a mass through interactions even when the bare Lagrangian is massless. This mechanism has been extensively investigated in QED$_4$~\cite{johnson1964self,10.1143/PTP.54.860,Fukuda1976,Curtis1990,Atkinson1993}, QED$_3$~\cite{Appelquist1981,appelquist1988critical,nash1989higher,maris1995confinement,Dagotto1989,Curtis1992,Bashir1994}, and PQED~\cite{alves2013chiral,fernandez2021dynamical,alves2017dynamical}. However, in PPQED the role of the screening scale parameter $m$ in determining critical quantities such as the critical coupling $\alpha_c$ and the critical flavor number $N_c$ has not been systematically clarified. Understanding this dependence is important because it reveals how a massive gauge sector reshapes the critical behavior relative to both PQED and QED$_3$.

In this work, we study dynamical mass generation in PPQED through Schwinger–Dyson equations~\cite{dyson1949s,schwinger1951green} in rainbow-quenched and unquenched truncations. Our main goal is to identify the qualitative and semi-quantitative impact of the pseudo-Proca screening scale on chiral symmetry breaking, rather than to extract precision values for the critical parameters. More specifically, we examine how the $(3+1)$D gauge-sector mass parameter $m$, together with the other parameters of the model, modifies the critical behavior in comparison with PQED and QED$_3$. We then assess the stability of this picture by incorporating a Ball–Chiu vertex construction beyond the bare rainbow approximation. In addition, we extend the analysis to an anisotropic regime with Fermi velocity $v_F\neq c$, with the aim of capturing the leading trend expected in two-dimensional Dirac materials such as graphene, where $v_F \approx c/300$.

From a conceptual viewpoint, it is important to stress that the parameter $m$ in PPQED should be interpreted as a screening scale inherited from the higher-dimensional massive gauge sector, rather than as a physical photon mass in $(2+1)$D or as an ultraviolet cutoff for the reduced theory. Its role is to control the crossover from long-range Coulomb behavior to Yukawa-screened interactions within the non-local planar theory. This interpretation is consistent with the monotonic suppression of dynamical mass generation as $m$ increases found throughout our analysis.

The paper is organized as follows. In Sec.~\ref{sec:Model}, we introduce the PPQED framework and derive the effective planar model from the higher-dimensional Proca-Stueckelberg theory. In Sec.~\ref{sec:SDequation}, we present the Schwinger–Dyson equations for the fermion and gauge sectors. Secs.~\ref{sec:RQApprox} and~\ref{sec:RUnQApprox} are devoted to the rainbow-quenched and rainbow-unquenched analyses, respectively. In Sec.~\ref{sec:BallChiu}, we go beyond the bare rainbow approximation by incorporating the Ball–Chiu vertex. In Sec.~\ref{sec:Anisotropy}, we discuss the anisotropic extension of the model. Finally, in Sec.~\ref{sec:Conclusion}, we summarize our results and comment on possible directions for future work. Additional numerical details are presented in~\ref{A} and~\ref{B}, while~\ref{C} contains the derivation of the corrected static potential.

\section{The model and its Feynman rules}\label{sec:Model}

In this section, we introduce the PPQED framework and derive its effective planar form from a higher-dimensional gauge-invariant massive theory. Our purpose is twofold. First, we wish to clarify the origin of the non-local gauge sector that characterizes PPQED. Second, we establish the Feynman rules that will be used in the Schwinger--Dyson analysis developed in the following sections.

We begin with the Proca--Stueckelberg model in $(3+1)$D, which provides a gauge-invariant description of a massive spin-1 field. In contrast to the ordinary Proca theory, where the mass term explicitly breaks gauge invariance, the Stueckelberg construction restores gauge symmetry through the introduction of an auxiliary scalar field $\phi$. In Euclidean space, the corresponding Lagrangian density is
\begin{equation}
\mathcal{L}_{\mathrm{PS}}
=
\frac{1}{4} F_{\mu\nu}F^{\mu\nu}
+
\frac{m^{2}}{2}
\left(
A_{\mu}
-
\frac{1}{m}\partial_{\mu}\phi
\right)^{2}
+
\frac{\lambda}{2}
(\partial_{\mu}A^{\mu})^{2},
\label{eq:LPS}
\end{equation}
where $F_{\mu\nu}=\partial_{\mu}A_{\nu}-\partial_{\nu}A_{\mu}$, $\lambda$ is the gauge-fixing parameter, and we adopt natural units, $c=\hbar=1$. The action is invariant under the gauge transformations
\begin{equation}
A_{\mu}\rightarrow A_{\mu}+\partial_{\mu}\theta,
\qquad
\phi\rightarrow \phi + m\theta,
\end{equation}
for an arbitrary sufficiently regular scalar function $\theta$. In this way, the Stueckelberg field restores gauge invariance without removing the massive scale from the theory~\cite{stuckelberg1938interaction,proca1936theoriee}.

As a first step toward dimensional reduction, we integrate out the auxiliary scalar field $\phi$, thereby obtaining an effective gauge-field Lagrangian of the form
\begin{equation}
\mathcal{L}_{\mathrm{PS}}^{\mathrm{eff}}
=
\frac{1}{4}F_{\mu\nu}F^{\mu\nu}
+
\frac{m^{2}}{2}A_{\mu}A^{\mu}
+
\frac{\lambda_{\Box}}{2}
(\partial_{\mu}A^{\mu})^{2},
\label{eq:LPS_eff}
\end{equation}
where
\begin{equation}
\lambda_{\Box}=-\lambda + m^{2}\Box^{-1},
\end{equation}
and $\Box$ denotes the d'Alembertian operator. We then minimally couple the gauge sector to Dirac fermions in $(3+1)$D, which leads to
\begin{align}\label{eq:L_eff_4D}
    \mathcal{L}_{\mathrm{eff}} &=  \frac{1}{4} F_{\mu \nu} F^{\mu \nu} + \frac{m^2}{2} A_{\mu}A^{\mu} + \frac{\lambda_{\Box}}{2} (\partial_{\mu}A^{\mu})^2 \nonumber \\
    &+ \bar\psi(i \slashed{\partial} - M)\psi - e A_{\mu}J^{\mu},
\end{align}
where $\slashed{\partial} = \gamma^{\mu}\partial_{\mu}$, $e$ is the electric charge, $M$ is the fermion mass, and $J^{\mu}=\bar{\psi}\gamma^{\mu}\psi$ is the conserved matter current. Throughout this work, we employ a four-component fermionic representation, with $4\times4$ Dirac matrices, which is the standard framework for discussing chiral symmetry and dynamical mass generation in planar gauge theories~\cite{Appelquist1981,nash1989higher,pisarski1984chiral,Kondo1990,Gusynin1996}.

Integrating out the gauge field $A_{\mu}$, we obtain the effective current--current interaction
\begin{equation}
S_{\mathrm{PS}}^{\mathrm{eff}}(J)
=
\int d^{4}x\, d^{4}y\,
\left[
\frac{1}{2}
J^{\mu}(x)\,
\Delta_{\mu\nu}(x-y)\,
J^{\nu}(y)
\right],
\label{eq:SeffJ}
\end{equation}
where the gauge-field propagator is given by
\begin{equation}
    \Delta_{\mu \nu} = \frac{1}{-\Box + m^2} \left( \delta_{\mu \nu} - \frac{\lambda_{\Box}}{-\Box + m^2-\lambda_{\Box}} \partial_{\mu} \partial_{\nu} \right).
    \label{eq:Delta_general}
\end{equation}

For a conserved current, $\partial_{\mu}J^{\mu}=0$, all gauge-dependent longitudinal contributions vanish, and only the transverse part of the propagator contributes. Therefore, in coordinate space,
\begin{equation}
\Delta_{\mu\nu}(x-y)
=
\int \frac{d^{4}k}{(2\pi)^{4}}
\,e^{-ik\cdot(x-y)}
\frac{\delta_{\mu\nu}}{k^{2}+m^{2}}.
\label{eq:Delta_simplified}
\end{equation}
This expression makes it clear that the interaction mediated by the massive gauge field is of Yukawa type. Indeed, the static potential follows from the Fourier transform of $\Delta_{00}(k_{0}=0,\mathbf{k})$, namely,
\begin{equation}
V(r)
=
\int \frac{d^{3}k}{(2\pi)^{3}}
\frac{e^{-i\mathbf{k}\cdot\mathbf{r}}}{k^{2}+m^{2}}
=
\frac{e^{-mr}}{4\pi r},
\label{eq:Yukawa_4D}
\end{equation}
where $r = \abs{\mathbf{r}} = \abs{\mathbf{x}-\mathbf{y}}$, which is the standard Yukawa potential~\cite{yukawa1935interaction}.

We now construct the reduced planar theory by following the same dimensional-reduction strategy employed in PQED~\cite{marino1993quantum}. To this end, we assume that the matter current is confined to the plane $x_{3}=0$, while the gauge field still propagates in the ambient $(3+1)$-dimensional space. Accordingly, we write
\begin{equation}
J^{\mu}(x_{0},x_{1},x_{2},x_{3})
=
\begin{cases}
j^{\mu}(x_{0},x_{1},x_{2})\,\delta(x_{3}), & \mu=0,1,2,\\[4pt]
0, & \mu=3.
\end{cases}
\label{eq:current_projection}
\end{equation}
Substituting this constraint into Eq.~\eqref{eq:SeffJ}, the effective action can be expressed entirely in terms of the planar current $j^{\mu}$~\cite{alves2018two}
\begin{equation}
S_{3D}^{\mathrm{eff}}(j)
=
\int d^{3}x\, d^{3}y\,
\left[
\frac{1}{2}
j^{\mu}(x)\,
G_{\mu\nu}(x-y)\,
j^{\nu}(y)
\right],
\label{eq:Seff3D}
\end{equation}
where
\begin{equation}
G_{\mu\nu}(x-y)
=
\Delta_{\mu\nu}(x-y,x_{3}=0,y_{3}=0)
\end{equation}
is the effective gauge propagator in the reduced $(2+1)$-dimensional theory.

Using Eq.~\eqref{eq:Delta_simplified}, one finds
\begin{equation}
G_{\mu\nu}(x-y)
=
\int \frac{d^{3}k}{(2\pi)^{3}}
e^{-ik\cdot(x-y)}
\int \frac{dk_{3}}{2\pi}
\frac{\delta_{\mu\nu}}{k^{2}+k_{3}^{2}+m^{2}}.
\label{eq:Gmunu_intermediate}
\end{equation}
After performing the integration over $k_{3}$, the result becomes
\begin{equation}
G_{\mu\nu}(x-y)
=
\int \frac{d^{3}k}{(2\pi)^{3}}
e^{-ik\cdot(x-y)}
\frac{\delta_{\mu\nu}}{2\sqrt{k^{2}+m^{2}}},
\label{eq:Gmunu_final}
\end{equation}
which is the non-local propagator that defines the planar pseudo-Proca theory.

As expected, the static interaction derived from the reduced model reproduces the same Yukawa potential obtained in the higher-dimensional description. Indeed,
\begin{equation}
V(r)
=
\int \frac{d^{2}k}{(2\pi)^{2}}
\frac{e^{-i\mathbf{k}\cdot\mathbf{r}}}{2\sqrt{k^{2}+m^{2}}}
=
\frac{e^{-mr}}{4\pi r}.
\label{eq:Yukawa_3D}
\end{equation}
This result shows that the dimensional reduction preserves the screened character of the interaction while encoding it in a genuinely non-local $(2+1)$D gauge kernel.

The planar effective theory associated with Eq.~\eqref{eq:Gmunu_final} can be written in terms of the pseudo-Proca Lagrangian density

\begin{align}
    \mathcal{L}_{PP} = & \; \frac{1}{4} F_{\mu \nu}  K[\Box] F^{\mu \nu} + \frac{\lambda}{4} A^{\mu} \partial_{\mu}  K[\Box] \partial_{\nu} A^{\nu} \nonumber \\
    &  + \bar{\psi}(i \slashed{\partial} - M)\psi - e A_{\mu} j^{\mu},
    \label{eq:LPP}
\end{align}
where the pseudo-differential operator is~\cite{alves2018two,nascimento2024effective}

\begin{equation}
 K[\Box] = \frac{2\sqrt{-\Box+m^2}}{-\Box}=\int \frac{d^3k}{(2\pi)^3}\, e^{ikx}\,
\frac{2\sqrt{k^2+m^2}}{k^2}.   
\end{equation}

This operator is responsible for the non-locality of the reduced gauge sector and ensures that the interaction interpolates continuously between the long-range Coulomb regime and the screened Yukawa regime.

At this stage, it is important to clarify the physical meaning of the parameter $m$. In the original $(3+1)$D Proca theory, $m$ appears as the mass scale of the massive gauge mode. After dimensional reduction, however, the planar propagator no longer exhibits a genuine massive pole; instead, it acquires the non-local form of Eq.~\eqref{eq:Gmunu_final}, characterized by a branch-cut structure. Therefore, in PPQED the parameter $m$ should not be interpreted as a physical photon mass in $(2+1)$D, nor as an ultraviolet cutoff for the reduced theory. Rather, it acts as a screening scale inherited from the higher-dimensional gauge sector, controlling the crossover from Coulomb-like behavior to Yukawa-screened interactions in the planar effective theory. In particular, the absence of an isolated pole in the planar propagator indicates that the reduced theory does not contain a localized massive gauge excitation in the usual $(2+1)$D sense. From this point on, the parameter $m$ will be referred to as the pseudo-Proca screening parameter (or screening scale) of the effective $(2+1)$D theory, rather than as a physical gauge-boson mass.

The Feynman rules follow directly from Eq.~\eqref{eq:LPP}. In Euclidean space, the interaction factor at the fermion--gauge vertex is $-e\gamma_\mu$, whereas the bare fermion propagator is
\begin{equation}
    \vcenter{\hbox{
        \feynmandiagram [horizontal=a to b, baseline=(a.base)] {
            a -- [fermion, with arrow=0.5] b,
        };
    }}
    = S_F^{(0)}(p) =  \left(- \slashed{p} + M \right)^{-1}.
    \label{eq:fermion_propagator_bare}
\end{equation}
\vspace{0.2cm}
The bare gauge-field propagator obtained from Eq.~\eqref{eq:LPP} reads, in momentum space and in the Landau gauge limit $\lambda\to\infty$,
\begin{equation}
    \vcenter{\hbox{
        \feynmandiagram [horizontal=a to b, baseline=(a.base)] {
            a -- [photon, with arrow=0.5, blue] b,
        };
    }}
    =  \Delta_{\mu \nu}^{(0)} (p) =  \frac{1}{2 \sqrt{p^2+m^2}} \, P_{\mu \nu},
    \label{eq:gauge_propagator_bare}
\end{equation}
where $P_{\mu \nu} = \left(\delta_{\mu \nu} - \frac{p_{\mu} p_{\nu}}{p^2}\right)$ is the transverse projection operator.
This expression makes explicit how the pseudo-Proca screening scale modifies the gauge sector while preserving the transverse structure characteristic of a gauge-invariant formulation.

In the next section, we formulate the Schwinger--Dyson equations for the fermion and gauge propagators, which will provide the basis for our analysis of dynamical mass generation and critical behavior in PPQED.

\section{Schwinger--Dyson equations}
\label{sec:SDequation}

With the effective planar pseudo-Proca model and its Feynman rules in place, we now turn to its non-perturbative analysis through the Schwinger--Dyson equations (SDEs)~\cite{dyson1949s,schwinger1951green}. These equations provide an exact hierarchy relating the full Green functions of the theory to their bare counterparts, thereby resumming quantum corrections to all orders. In gauge theories, the SDE framework has long been recognized as a powerful tool for investigating chiral symmetry breaking, confinement, and vacuum structure, both in four-dimensional theories and in lower-dimensional analogues such as QED$_3$~\cite{roberts1994dyson,alkofer2001infrared,Appelquist1986}.

In the present context, our main interest is the onset of dynamical mass generation in PPQED. For this reason, we focus on the coupled SDEs for the fermion and gauge-field propagators. The full fermion propagator satisfies
\begin{equation}
S^{-1}_{F}(p)
=
\left(S^{(0)}_{F}(p)\right)^{-1}
-
\Xi(p),
\label{eq:SDE_fermion}
\end{equation}
where $\Xi(p)$ denotes the fermion self-energy. Likewise, the full pseudo-Proca propagator obeys
\begin{equation}
\left(\Delta_{\mu\nu}(p)\right)^{-1}
=
\left(\Delta_{\mu\nu}^{(0)}(p)\right)^{-1}
-
\Pi_{\mu\nu}(p),
\label{eq:SDE_gauge}
\end{equation}
where $\Pi_{\mu\nu}(p)$ is the gauge-field self-energy (vacuum polarization tensor). In diagrammatic language, these equations are represented in Figs.~\ref{fig:SDfermion} and~\ref{fig:SDphoton}, respectively.

\begin{figure}[H]
    \centering
    \begin{equation*}
        \begin{split}
            \vcenter{\hbox{
                $\left(
                \begin{tikzpicture}[baseline=(current bounding box.center)]
                \begin{feynman}
                    \vertex (a) at (0,0);
                    \vertex (b) at (0.7,0);
                    \vertex (c) at (1.4,0);
                    \diagram* {
                        (a) -- [fermion, black,  edge label=\(p\)] (b),
                        (b) -- [fermion, black,  edge label=\(p\)] (c),
                    };
                \end{feynman}
                \filldraw[fill=white] (b) circle (0.2cm);
                \filldraw[fill=gray!30, draw=black] (b) circle (0.2cm);
                \filldraw[pattern=north east lines, pattern color=black] (b) circle (0.2cm);
                \end{tikzpicture}
                \right)^{-1}$
            }}
            \;&=\;
            \vcenter{\hbox{
                $\left(
                \begin{tikzpicture}[baseline=(current bounding box.center)]
                \begin{feynman}
                    \vertex (a) at (0,0);
                    \vertex (b) at (0.9,0);
                    \diagram* {
                        (a) -- [fermion, black,  edge label=\(p\)] (b),          
                    };
                \end{feynman}
                \end{tikzpicture}
                \right)^{-1}$
            }} \\
            &\quad -\;
            \vcenter{\hbox{
                \begin{tikzpicture}[baseline=(current bounding box.center)]
                \begin{feynman}
                    \vertex (a) at (0,0);
                    \vertex (b) at (0.8,0);
                    \vertex (c) at (1.6,0.8);
                    \vertex (d) at (1.6,0);
                    \vertex (e) at (2.4,0);
                    \vertex (f) at (3.4,0);
                    \diagram* {
                        (a) -- [fermion, black, edge label=\(p\)] (b),
                        (b) -- [photon, blue, quarter left, looseness=1] (c),
                        (c) -- [photon, blue, quarter left, looseness=1] (e),
                        (b) -- [plain, black] (d),
                        (d) -- [plain, black] (e),
                        (e) -- [fermion, black, edge label=\(p\)] (f),
                    };
                \end{feynman}
                \draw[black, thick] (0.6, -0.4) -- (0.6, 0.4);
                \draw[black, thick] (2.7, -0.4) -- (2.7, 0.4);
                \filldraw[fill=white] (d) circle (0.2cm); 
                \filldraw[fill=gray!30, draw=black] (d) circle (0.2cm); 
                \filldraw[pattern=north east lines, pattern color=black] (d) circle (0.2cm);
                \node at (d) [below=0.2cm] {\(k\)};
                \filldraw[fill=black!70, draw=black] (c) circle (0.2cm);
                \node at (c) [above=0.2cm] {\(p-k\)};
                \filldraw[fill=gray!30, draw=black] (e) circle (0.2cm);
                \end{tikzpicture}
            }}
        \end{split}
    \end{equation*}
    \caption{The SDE for the full electron propagator. Filled dots indicate full propagators and vertex. The second term on the right-hand side represents the fermion self-energy $\Xi(p)$, given in Eq.~\eqref{eq:fermion_selfenergy}.}
    \label{fig:SDfermion}
\end{figure}

\begin{figure}[H]
    \centering
    \begin{equation*}
        \begin{split}
            \vcenter{\hbox{
                $\left(
                \begin{tikzpicture}[baseline=(current bounding box.center)]
                \begin{feynman}
                    \vertex (a) at (0,0);
                    \vertex (b) at (0.7,0);
                    \vertex (c) at (1.4,0);
                    \diagram* {
                        (a) -- [photon, blue,with arrow=0.5,  edge label=\(p\)] (b),
                        (b) -- [photon, blue, with arrow=0.5, edge label=\(p\)] (c),
                    };
                \end{feynman}
                \filldraw[fill=white] (b) circle (0.2cm); 
                \filldraw[fill=black!70, draw=black] (b) circle (0.2cm); 
                \end{tikzpicture}
                \right)^{-1}$
            }}
            \;&=\;
            \vcenter{\hbox{
                $\left(
                \begin{tikzpicture}[baseline=(current bounding box.center)]
                \begin{feynman}
                    \vertex (a) at (0,0);
                    \vertex (b) at (0.9,0);
                    \diagram* {
                        (a) -- [photon, blue, with arrow=0.5,  edge label=\(p\)] (b),          
                    };
                \end{feynman}
                \end{tikzpicture}
                \right)^{-1}$
            }} \\
            &\quad -\;
            \vcenter{\hbox{
                \begin{tikzpicture}[baseline=(current bounding box.center)]
                \begin{feynman}
                    \vertex (a) at (0,0);
                    \vertex (b) at (1.0,0);
                    \vertex (c) at (1.5,0.5);
                    \vertex (d) at (1.5,-0.5);
                    \vertex (e) at (2.0,0);
                    \vertex (f) at (3.3,0);
                    \diagram* {
                        (a) -- [photon, blue, with arrow=0.5, edge label=\(p\)] (b),
                        (b) -- [plain, quarter left, looseness=1] (c),
                        (d) -- [plain, quarter left, looseness=1] (b),
                        (c) -- [plain, quarter left, looseness=1] (e),
                        (e) -- [plain, quarter left, looseness=1] (d),
                        (e) -- [photon, blue, with arrow=0.5, edge label=\(p\)] (f),
                    };
                \end{feynman}
                \draw[black, thick] (0.7, -0.4) -- (0.7, 0.4);
                \draw[black, thick] (2.45, -0.4) -- (2.45, 0.4);
                \filldraw[fill=white] (d) circle (0.2cm);
                \filldraw[fill=gray!30, draw=black] (d) circle (0.2cm);
                \filldraw[pattern=north east lines, pattern color=black] (d) circle (0.2cm);
                \node at (d) [below=0.2cm] {\(k\)};
                \filldraw[fill=white] (c) circle (0.2cm); 
                \filldraw[fill=gray!30, draw=black] (c) circle (0.2cm); 
                \filldraw[pattern=north east lines, pattern color=black] (c) circle (0.2cm);
                \node at (c) [above=0.2cm] {\(p+k\)};
                \filldraw[fill=gray!30, draw=black] (e) circle (0.2cm);
                \end{tikzpicture}
            }}
        \end{split}
    \end{equation*}
    \caption{The SDE for the full gauge-field propagator. Filled dots indicate full propagators and vertex. The second term on the right-hand side represents the gauge-field self-energy $\Pi^{\mu \nu}(p)$, given by Eq.~\eqref{eq:gauge_selfenergy}.}
    \label{fig:SDphoton}
\end{figure}

The corresponding self-energy functions are given by
\begin{equation}
\Xi(p)
=
\int \frac{d^{3}k}{(2\pi)^{3}}
\,
(-e\gamma^{\mu})\,
S_{F}(k)\,
(e  \Gamma^{\nu}(k,p))\,
\Delta_{\mu\nu}(p-k),
\label{eq:fermion_selfenergy}
\end{equation}
and

\begin{equation}
\begin{split}
\Pi^{\mu \nu}(p) &= - \int \frac{d^3 k}{(2 \pi)^3}\,
\Tr\Big[ (-e\gamma^{\mu}) S_F(p+k) \\
&\qquad\qquad \times (e\Gamma^{\nu}(k,p))\, S_F(k) \Big],
\end{split}
\label{eq:gauge_selfenergy}
\end{equation}
where $S_{F}(k)$ is the full fermion propagator, $\Gamma_{\nu}(k,p)$ is the full fermion--gauge vertex, and $\Delta_{\mu\nu}(p-k)$ is the dressed pseudo-Proca propagator.

At this stage, the theory is still exact. However, as in other strongly coupled gauge theories, the full SDE hierarchy is infinite, since the propagator equations are coupled to the vertex equation, and the latter is in turn coupled to higher $n$-point functions. In practice, one must therefore introduce a truncation scheme in order to obtain a closed and tractable system. The main issue is then to choose a truncation that is sufficiently simple to allow analytical and numerical progress, while still preserving the relevant physical content of the problem.

The interaction vertex itself satisfies a Schwinger--Dyson equation~\cite{roberts1994dyson}. In the present work, however, we do not attempt to solve the full vertex equation exactly. Instead, we employ two complementary levels of approximation. In the first one, we adopt the standard rainbow truncation, in which the full vertex is replaced by its lowest-order form. In the second one, we consider the Ball--Chiu vertex, which provides a non-perturbative ansatz consistent with gauge invariance and free of kinematic singularities~\cite{BallChiu1980}. In particular, the Ball--Chiu construction satisfies the Ward--Takahashi identity
\begin{equation}
(p-k)_{\mu}\Gamma^{\mu}(p,k)
=
S^{-1}_{F}(p)-S^{-1}_{F}(k),
\label{eq:WTI}
\end{equation}
thereby relating the vertex directly to the structure of the full fermion propagator.

Motivated by the standard decomposition used in QED$_4$, QED$_3$, and PQED~\cite{alves2013chiral,alves2017dynamical,roberts1994dyson,Bashir_2012}, we parametrize the full fermion propagator as
\begin{equation}
S^{-1}_{F}(p)
=
-\slashed{p}\,A(p)
+
\Sigma(p),
\label{eq:full_fermion_ansatz}
\end{equation}
where $A(p)$ is the wavefunction renormalization and $\Sigma(p)$ is the mass function. Dynamical mass generation is signaled by the existence of a non-trivial solution for $\Sigma(p)$ even when the bare fermion mass is set to zero. Thus, the central object in our analysis will be the gap equation obtained from the fermion SDE under different truncation schemes.

Using the four-component Dirac representation and taking the trace of Eq.~\eqref{eq:full_fermion_ansatz} and the fermion Schwinger--Dyson equation~\footnote{In $(2+1)$D Euclidean space and in the $4\times 4$ representation, the Dirac traces are given by $\operatorname{Tr}(\gamma^\mu \gamma^\nu) = -4 \delta^{\mu\nu}$, $\operatorname{Tr}(\gamma^\mu \gamma^\nu \gamma^\alpha) = 0$, and $\operatorname{Tr}(\gamma^\mu \gamma^\nu \gamma^\alpha \gamma^\beta) = 4(\delta^{\mu\nu} \delta^{\alpha\beta} - \delta^{\mu\alpha} \delta^{\nu\beta} + \delta^{\mu\beta} \delta^{\nu\alpha})$.}, Eq.~\eqref{eq:SDE_fermion}, we obtain the gap equations
\begin{align}
    \Sigma(p) &= M + \frac{e^2}{\Tr{\mathbb{I}}} \int \frac{d^3 k}{(2 \pi)^3} \nonumber \\
    &\qquad\times \Tr\left\{\gamma^{\mu} S_F(k) \Gamma^{\nu}(k,p)\right\}  \Delta_{\mu \nu}(p-k), \label{eq:sigma_initial}\\[0.3cm]
    A(p) &= 1 +  \frac{e^2}{p^2 \Tr(\mathbb{I})} \int \frac{d^3 k}{(2 \pi)^3} \nonumber \\
    &\qquad \times \Tr\left\{\slashed{p} \   \gamma^{\mu} S_F(k) \Gamma^{\nu}(k,p) \right\} \Delta_{\mu \nu}(p-k). \label{eq:A_initial}
\end{align}

It is worth highlighting that our strategy is not to extract precision values for all critical quantities from a fully self-consistent solution of the complete SDE tower. Rather, our goal is to identify the qualitative and semi-quantitative role of the pseudo-Proca screening scale in the chiral dynamics of the theory. 

For this reason, we begin with the simplest truncation, which already captures the leading structure of the phase transition and allows analytical insight into the dependence of the critical thresholds on the screening parameter $m$.

In the following sections, we distinguish between results obtained within the rainbow truncation and those that remain stable under vertex-improved treatments. This separation is important because the rainbow approximation is expected to capture the leading structure of the phase transition, whereas the improved vertices mainly shift the numerical values of the critical parameters.

In what follows, we work in the chiral limit $M=0$ and investigate whether a non-trivial fermion mass function is generated dynamically.

We therefore proceed in stages: we first analyze the rainbow-quenched approximation, then include vacuum-polarization effects in the unquenched case, and finally test the robustness of the resulting physical picture through a Ball--Chiu vertex construction. Our purpose is to identify the qualitative and semi-quantitative role of the pseudo-Proca screening scale in the chiral dynamics of the theory, while keeping track of how the resulting critical thresholds depend on the truncation scheme.

\section{Rainbow-quenched approximation}
\label{sec:RQApprox}

As a first step in the non-perturbative analysis, we consider the rainbow-quenched truncation. In this approximation, the full fermion--gauge vertex is replaced by the bare Dirac vertex,
\begin{equation}
\Gamma^{\mu}(k,p)\to -\gamma^{\mu},
\end{equation}
while the dressed pseudo-Proca propagator is replaced by its bare form,
\begin{equation}
\Delta_{\mu\nu}(p)\to \Delta^{(0)}_{\mu\nu}(p),
\end{equation}
as is standard in related Schwinger--Dyson studies~\cite{alkofer2001infrared,maris2003dyson}. This truncation provides the simplest setting in which the role of the pseudo-Proca screening scale can be analyzed analytically.

In the chiral limit, the existence of a non-trivial solution for the mass function $\Sigma(p)$ is sufficient to characterize a phase with dynamical chiral symmetry breaking. In the present section, and for the sake of analytical transparency, we also adopt the approximation $A(p)\approx 1$, which will be relaxed later in Sec.~\ref{sec:BallChiu}. This first-order truncation is intended to isolate the effect of the pseudo-Proca screening scale on the critical structure of the theory. 

Applying these approximations in Eq.~\eqref{eq:sigma_initial} results in
\begin{equation}
\Sigma(p)
=
e^{2}
\int \frac{d^{3}k}{(2\pi)^{3}}
\frac{\Sigma(k)\,\delta_{\mu\nu}}{k^{2}+\Sigma^{2}(k)}
\Delta_{\mu\nu}^{(0)}(p-k).
\label{eq:gap_quenched_1}
\end{equation}

Substituting the bare pseudo-Proca propagator from Eq.~\eqref{eq:gauge_propagator_bare}, introducing the fine-structure constant $\alpha=e^{2}/4\pi$, and performing the angular integration, Eq.~\eqref{eq:gap_quenched_1} can be written as
\begin{equation}
\Sigma(p)
=
\frac{2\alpha}{\pi}
\int_{0}^{\Lambda}
dk\,
\frac{k^{2}\Sigma(k)}{k^{2}+\Sigma^{2}(k)}
K(k,p),
\label{eq:gap_quenched_2}
\end{equation}
where $\Lambda$ is the ultraviolet cutoff and the kernel is given by
\begin{equation}
K(k,p)
=
\frac{\sqrt{(k+p)^{2}+m^{2}}-\sqrt{(k-p)^{2}+m^{2}}}{2kp}.
\label{eq:kernel_exact}
\end{equation}

The structure of Eq.~\eqref{eq:gap_quenched_2} already makes the role of the pseudo-Proca screening parameter transparent: increasing $m$ suppresses the kernel and therefore weakens the effective interaction responsible for gap formation. In order to extract analytical scaling laws, we now approximate the kernel by its leading contribution in each momentum regime~\cite{Appelquist1981}, namely,
\begin{equation}
K(k,p)
=
\frac{\Theta(p-k)}{\sqrt{p^{2}+m^{2}}}
+
\frac{\Theta(k-p)}{\sqrt{k^{2}+m^{2}}},
\label{eq:kernel_approx}
\end{equation}
where $\Theta(x)$ is the Heaviside step function. This approximation is introduced only to obtain a closed differential form for the gap equation; the full integral equation remains the more fundamental starting point for the numerical analysis.

With Eq.~\eqref{eq:kernel_approx}, the integral equation becomes

\begin{align}
    \Sigma(p) &= \frac{2  \alpha}{\pi}  \int_0^{p}  dk   \frac{ k^2 \Sigma(k)  }{k^2 + \Sigma^2(k)} \frac{1}{(p^2+m^2)^{1/2}}  \nonumber \\ 
    &\quad+ \frac{ 2 \alpha}{\pi}  \int_p^{\Lambda}  dk   \frac{ k^2 \Sigma(k)  }{k^2 + \Sigma^2(k)} \frac{1}{(k^2+m^2)^{1/2}}.
    \label{eq:gap_quenched_split}
\end{align}
Differentiating Eq.~\eqref{eq:gap_quenched_split} with respect to $p$ and then differentiating once more, one obtains the second-order differential equation
\begin{equation}
\frac{d}{dp}
\left[
\frac{(p^{2}+m^{2})^{3/2}}{p}\,
\Sigma'(p)
\right]
+
\frac{2 \alpha}{\pi}
\frac{p^{2}\Sigma(p)}{p^{2}+\Sigma^{2}(p)}
=
0.
\label{eq:gap_diff_exact}
\end{equation}

The corresponding infrared and ultraviolet boundary conditions follow directly from Eq.~\eqref{eq:gap_quenched_split}. In the ultraviolet region, one finds
\begin{equation}
\lim_{p\to\Lambda}
\left[
\frac{p^{2}+m^{2}}{p}\Sigma'(p)+\Sigma(p)
\right]
=
0,
\label{eq:BC_UV}
\end{equation}
whereas in the infrared region,
\begin{equation}
\lim_{p\to 0}
\left[
\frac{(p^{2}+m^{2})^{3/2}}{p}\Sigma'(p)
\right]
=
0.
\label{eq:BC_IR}
\end{equation}

Eq.~\eqref{eq:gap_diff_exact} does not admit a simple closed-form solution in its full non-linear form. Nevertheless, analytical progress can be made in the regime $p^{2}\gg \Sigma^{2}(p)$, where the equation may be linearized. Since our purpose at this stage is to obtain the asymptotic scaling behavior and an analytical estimate for the critical coupling, we further approximate
\begin{equation}
p^{2}\left(1+\frac{m^{2}}{p^{2}}\right)^{3/2}
\approx
p^{2}\left(1+\frac{m^{2}}{\Lambda^{2}}\right)^{3/2},
\label{eq:UV_approx}
\end{equation}
which is consistent with the ultraviolet-dominated regime of the integral equation when $m\ll \Lambda$~\cite{Burgess2007}. The consistency of this approximation is checked a posteriori by comparison with the numerical solution of the full integral equation (see~\ref{A}). Under this approximation, Eq.~\eqref{eq:gap_diff_exact} reduces to
\begin{equation}
p^{2}\Sigma''(p)
+
2p\Sigma'(p)
+
\frac{2\alpha}{\pi\left(1+\frac{m^{2}}{\Lambda^{2}}\right)^{3/2}}
\Sigma(p)
=
0.
\label{eq:gap_diff_linear}
\end{equation}
This linearized equation is intended to determine the critical bifurcation structure of the theory in the ultraviolet-dominated regime. Accordingly, the resulting expression for the critical coupling should be interpreted as a controlled asymptotic estimate rather than as an exact critical value. It does not describe the full infrared saturation of the mass function, which is why the analytical discussion is complemented by the numerical solutions presented in~\ref{A}. The general solution of Eq.~\eqref{eq:gap_diff_linear} may be written as

\begin{equation}
\Sigma(p)
=
A_{1}\,p^{-\lambda_{1}/2}
+
A_{2}\,p^{\lambda_{1}/2-1},
\label{eq:gap_power_solution}
\end{equation}
where the exponent is conveniently parametrized as
\begin{equation}
\lambda_{1}
=
1-\sqrt{1-\frac{\alpha}{\alpha_{c}}}.
\label{eq:lambda1}
\end{equation}
From this expression one identifies the critical fine-structure constant
\begin{equation}
\alpha_{c}(m,\Lambda)
=
\frac{\pi}{8}
\left(
1+\frac{m^{2}}{\Lambda^{2}}
\right)^{3/2}.
\label{eq:alpha_c_quenched}
\end{equation}
Eq.~\eqref{eq:alpha_c_quenched} shows explicitly that increasing the pseudo-Proca screening parameter raises the threshold for chiral symmetry breaking. In other words, the Yukawa screening induced by $m$ suppresses dynamical mass generation, so that a larger coupling is required to sustain a non-trivial fermion gap.

In the continuum limit $\Lambda\to\infty$, Eq.~\eqref{eq:gap_power_solution} satisfies both IR and UV boundary conditions for any value of $\alpha$, yielding dynamical mass generation regardless of the coupling constant. This result is consistent with QED$_4$~\cite{atkinson1987bifurcation} and PQED~\cite{alves2013chiral}.

For the case of finite $\Lambda$ and within the linearized ultraviolet analysis, the regime $\alpha<\alpha_c$ leads to non-oscillatory solutions, and the boundary conditions select the trivial solution for the mass function. By contrast, when $\alpha>\alpha_{c}$ the exponent becomes complex, and the physically relevant solution takes the oscillatory form
\begin{equation}
\Sigma(p)
=
\frac{C}{\sqrt{p}}
\sin\left[
\beta
\left(
\ln\frac{p}{\Sigma_{0}}+\delta
\right)
\right],
\label{eq:gap_oscillatory}
\end{equation}
where
\begin{equation}
2\beta
=
\sqrt{\frac{\alpha}{\alpha_{c}}-1},
\label{eq:beta_def}
\end{equation}
and $\delta$ is a phase fixed by the boundary conditions. Imposing the ultraviolet condition, Eq.~\eqref{eq:BC_UV}, one obtains the scaling factor
\begin{equation}
\Sigma_{0}
=
\Lambda
\exp\left[
\delta
+
\frac{2(m^{2}+\Lambda^{2})}{\Lambda^{2}-m^{2}}
-
\frac{\pi n}{\beta}
\right],
\label{eq:Miransky_scale}
\end{equation}
with integer $n$ labeling the branches of the solution.

Eq.~\eqref{eq:Miransky_scale} exhibits the characteristic Miransky-type scaling of strongly coupled gauge theories, usually discussed for both QED$_3$ and QED$_4$~\cite{Fomin1983,Miransky1985A,Miransky1985B}. In particular, the non-analytic dependence on $(\alpha - \alpha_c)$ signals a conformal phase transition~\cite{Kaplan2009,Miransky1997}. This expression is to be interpreted within the regime $m<\Lambda$, and in practice within the hierarchy $m\ll\Lambda$ assumed throughout the analytical treatment.

The present quenched analysis therefore provides a clear first picture of the role played by the pseudo-Proca screening scale: increasing $m$ weakens the kernel, raises the critical coupling, and suppresses the dynamically generated fermion mass. In the limit $m \to 0$, both Eq.~\eqref{eq:gap_power_solution} and Eq.~\eqref{eq:Miransky_scale} are in agreement with the corresponding results of PQED~\cite{alves2013chiral}.

Building on these results obtained in the simplest truncation, we now improve the analysis by incorporating vacuum-polarization effects in the gauge sector through the rainbow-unquenched approximation. This requires extending the model to $N$ fermion flavors, which naturally leads to the notion of a critical flavor number.

\section{Rainbow-unquenched approximation}
\label{sec:RUnQApprox}

We now improve the quenched analysis by incorporating vacuum-polarization effects in the gauge sector. This is the natural next step in the Schwinger--Dyson hierarchy, since it allows the pseudo-Proca propagator to respond dynamically to fermionic fluctuations while still keeping the fermion--gauge vertex at the bare level. In practice, this corresponds to the rainbow-unquenched approximation, which is most conveniently implemented within the large-$N$ expansion~\cite{Appelquist1981,Curtis1992,roberts1994dyson}.

To this end, we extend the model to $N$ fermion flavors,
\begin{align}
    \mathcal{L}_{PP} &= \frac{1}{4} F_{\mu \nu} K[\Box] F^{\mu \nu} + \frac{\lambda}{4} A^{\mu} \partial_{\mu} K[\Box] \partial_{\nu} A^{\nu}  \nonumber \\ 
    &+ \sum_{a=1}^{N}\bar{\psi}_a(i \slashed{\partial}- e \gamma^{\mu}A_{\mu}) \psi_a,
    \label{eq:LPP_N}
\end{align}
where $a = 1, \dots, N$ is the flavor index. As usual in the large-$N$ framework, we keep the combination $4\pi\alpha \equiv \frac{g}{N}$ fixed as $N$ is varied. In this notation, $g$ plays the role of the effective interaction parameter that controls the strength of the vacuum-polarization correction. This convention will be used throughout the unquenched analysis so that the corresponding critical behavior can be compared directly with the quenched results expressed in terms of $\alpha$. At leading order in $1/N$, the gauge-field self-energy is determined by the one-loop polarization tensor of massless Dirac fermions.

In Landau gauge, it has the standard transverse form

\begin{equation}
\Pi_{\mu\nu}(p)
=
\Pi(p^{2})
\left(
\delta_{\mu\nu}
-
\frac{p_{\mu}p_{\nu}}{p^{2}}
\right),
\label{eq:Pi_tensor}
\end{equation}
with scalar part given by~\cite{nash1989higher,maris1995confinement,pisarski1984chiral,gusynin1995dynamical}
\begin{equation}
\Pi(p^{2})
=
-\frac{g}{8}\sqrt{p^{2}}.
\label{eq:Pi_scalar}
\end{equation}
Substituting Eq.~\eqref{eq:Pi_scalar} into the gauge Schwinger--Dyson equation, Eq.~\eqref{eq:SDE_gauge}, and using the bare pseudo-Proca propagator from Sec.~\ref{sec:Model}, one obtains the dressed gauge-field propagator
\begin{equation}
\Delta_{\mu\nu}(p)
=
\frac{1}{
2\sqrt{p^{2}+m^{2}}+\frac{g}{8}\sqrt{p^{2}}
}
\left(
\delta_{\mu\nu}
-
\frac{p_{\mu}p_{\nu}}{p^{2}}
\right).
\label{eq:Delta_unquenched}
\end{equation}
This expression makes explicit the interplay between the pseudo-Proca screening parameter and vacuum polarization: the denominator now contains both the mass-dependent non-local term and the fermion-induced screening contribution proportional to $g$. From Eq.~\eqref{eq:Delta_unquenched} we can calculate an explicit expression for the corrected static potential of PPQED, see~\ref{C}.

As in Sec.~\ref{sec:RQApprox}, we work in the chiral limit and focus on the mass function $\Sigma(p)$. Within the rainbow-unquenched truncation, we take
\begin{equation}
\Gamma^{\mu}(k,p)\to -\gamma^{\mu},
\qquad
A(p)\simeq 1,
\end{equation}
so that the fermion Schwinger--Dyson equation reduces to an integral equation for $\Sigma(p)$ alone. After applying these approximations in Eq.~\eqref{eq:sigma_initial} and contracting the transverse tensor structure of Eq.~\eqref{eq:Delta_unquenched}, we obtain
\begin{align}
    \Sigma(p)  &= \frac{g}{N} \int \frac{d^3 k}{(2 \pi)^3}  \frac{\Sigma(k)  }{ k^2 + \Sigma^2(k)  }    \nonumber \\  & \quad \times \frac{1}{ \sqrt{(p-k)^2+m^2} + \frac{g}{16} \sqrt{(p-k)^2}}.
\label{eq:gap_unquenched_integral}
\end{align}

Compared with the quenched case, the kernel is now additionally suppressed by the vacuum-polarization term. Physically, this means that the dressed gauge sector tends to further weaken the interaction responsible for chiral symmetry breaking.

Proceeding as in Sec.~\ref{sec:RQApprox}, we reduce the integral equation to a differential form that makes the critical behavior more transparent. After performing the angular integration and adopting the same ultraviolet-oriented treatment used previously, the gap equation may be cast as
\begin{equation}
\frac{d}{dp}
\left[
\mathcal{F}(m,p,g)\,p^{2}\Sigma'(p)
\right]
+
\frac{g}{2N\pi^{2}}
\frac{p^{2}\Sigma(p)}{p^{2}+\Sigma^{2}(p)}
=
0,
\label{eq:gap_unquenched_diff}
\end{equation}
where, for notational clarity, we define
\begin{equation}
\mathcal{F}(m,p,g)
=\sqrt{1 + \frac{m^2}{p^2}}\frac{ \left( \frac{g}{16} +  \sqrt{1 + \frac{m^2}{p^2}} \right)^2 }{ \left(  \frac{g}{16} \sqrt{1 + \frac{m^2}{p^2}} +1\right)}.
\label{eq:F_mpg}
\end{equation}
This function encodes the combined effect of the pseudo-Proca scale and the leading vacuum-polarization correction. The ultraviolet and infrared boundary conditions follow from the integral representation and take the form
\begin{equation}
\lim_{p\to\Lambda}
\left[
\mathcal{B}(m,p,g)\,p\,\Sigma'(p)+\Sigma(p)
\right]
=
0,
\label{eq:BC_UV_unquenched}
\end{equation}
and
\begin{equation}
\lim_{p\to 0}
\left[
\mathcal{F}(m,p,g)\,p^{2}\Sigma'(p)
\right]
=
0,
\label{eq:BC_IR_unquenched}
\end{equation}
where the ultraviolet boundary coefficient is given by
\begin{equation}
\mathcal{B}(m,p,g)
=
\sqrt{1+\frac{m^{2}}{p^{2}}}
\frac{
\sqrt{1+\frac{m^{2}}{p^{2}}}+\frac{g}{16}
}{
1+\frac{g}{16}\sqrt{1+\frac{m^{2}}{p^{2}}}
}.
\label{eq:B_mpg}
\end{equation}

Although Eq.~\eqref{eq:gap_unquenched_diff} remains non-linear, analytical progress can again be made in the regime $p^{2}\gg \Sigma^{2}(p)$, where the equation may be linearized. As in the quenched case, our purpose is to extract the asymptotic scaling structure and an analytical estimate of the critical threshold. We therefore approximate
\begin{equation}
\mathcal{F}(m,p,g)\approx \mathcal{F}(m,\Lambda,g),
\label{eq:F_UV_approx}
\end{equation}
which is appropriate in the ultraviolet-dominated regime of the gap equation. Under this approximation, Eq.~\eqref{eq:gap_unquenched_diff} reduces to
\begin{equation}
p^{2}\Sigma''(p)
+
2p\Sigma'(p)
+
\frac{g}{2N\pi^{2}}
\frac{1}{\mathcal{F}(m,\Lambda,g)}
\Sigma(p)
=
0.
\label{eq:gap_unquenched_linear}
\end{equation}
As before, this linearized equation should be understood as an analytical approximation designed to expose the critical structure of the theory; it does not replace the full integral equation as the appropriate framework for quantitative analysis.

The general solution of Eq.~\eqref{eq:gap_unquenched_linear} can be written as
\begin{equation}
\Sigma(p)
=
C_{1}\,p^{-\lambda_{2}/2}
+
C_{2}\,p^{\lambda_{2}/2-1},
\label{eq:Sigma_power_unquenched}
\end{equation}
where
\begin{equation}
\lambda_{2}
=
1-
\sqrt{
1-\frac{2g}{N\pi^{2}\mathcal{F}(m,\Lambda,g)}
}.
\label{eq:lambda2}
\end{equation}
From this expression and the corresponding boundary conditions, one identifies the critical number of fermion flavors,
\begin{equation}
N_{c}(m,\Lambda,g)
=
\frac{2g}{\pi^{2}\mathcal{F}(m,\Lambda,g)}.
\label{eq:Nc_unquenched}
\end{equation}
Therefore, within the present approximation, the system exhibits dynamical mass generation for $N<N_{c}(m,\Lambda,g)$, whereas for $N>N_{c}(m,\Lambda,g)$, the symmetric massless phase is selected by the boundary conditions.

Eq.~\eqref{eq:Nc_unquenched} provides a direct generalization of the familiar PQED and QED$_3$ results. In the limit $m\to 0$, we recover the PQED expression~\cite{alves2013chiral}
\begin{equation}
N_{c}(g)
=
\frac{32g}{\pi^{2}(g+16)}.
\label{eq:Nc_PQED_limit}
\end{equation}
Furthermore, in the strong-coupling limit $g\to \infty$, followed by $m\to 0$, one obtains
\begin{equation}
N_{c}
=
\frac{32}{\pi^{2}},
\label{eq:Nc_QED3_limit}
\end{equation}
which is the standard QED$_3$ result~\cite{appelquist1988critical}. In this sense, the pseudo-Proca theory interpolates smoothly between the PQED regime and the more strongly screened gauge dynamics characteristic of QED$_3$-like behavior.

When $N<N_{c}$, the exponent in Eq.~\eqref{eq:Sigma_power_unquenched} becomes complex, and the physically relevant solution takes the oscillatory form
\begin{equation}
\Sigma(p)
=
\frac{D}{\sqrt{p}}
\sin\left[
\gamma
\left(
\ln\frac{p}{\tilde{\Sigma}_{0}}
+
\tilde{\delta}
\right)
\right],
\label{eq:Sigma_osc_unquenched}
\end{equation}
where
\begin{equation}
2\gamma
=
\sqrt{\frac{N_{c}}{N}-1}.
\label{eq:gamma_def}
\end{equation}
The constants $D$ and $\tilde{\delta}$ are fixed by the linear combination of the independent solutions and by the boundary conditions. By imposing the ultraviolet condition, Eq.~\eqref{eq:BC_UV_unquenched}, one obtains the scaling parameter
\begin{equation}
\tilde{\Sigma}_0 = \Lambda \exp \left[ \tilde{\delta} + \frac{2\mathcal{B}(m, \Lambda, g)}{2 - \mathcal{B}(m, \Lambda, g)} - \frac{n \pi}{\gamma} \right],
\label{eq:Miransky_unquenched}
\end{equation}
where $\mathcal{B}(m, \Lambda, g)$ is the boundary coefficient defined in Eq.~\eqref{eq:B_mpg}. This expression ensures that the scale factor remains consistent with the quenched limit $g \to 0$, as the exponential argument correctly reduces to the form found in Eq.~\eqref{eq:Miransky_scale}.

Eq.~\eqref{eq:Miransky_unquenched} exhibits the same essential Miransky-type scaling found in other strongly coupled gauge theories. In particular, the non-analytic dependence of the infrared screening scale on the distance to the critical point signals a conformal phase transition. The physical picture is therefore fully consistent with the quenched analysis, but now refined by the dynamical screening of the gauge sector: increasing the pseudo-Proca scale suppresses the gap, while vacuum polarization introduces an additional tendency toward restoration of the symmetric phase.

\begin{figure}[t]
    \centering
    \includegraphics[width=0.47\textwidth]{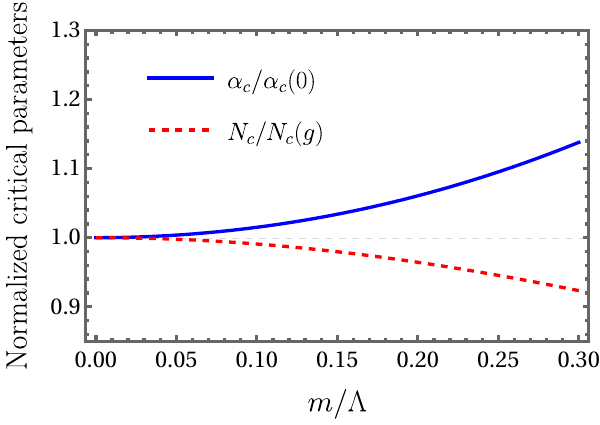}
    \caption{Dependence of the normalized critical coupling $\alpha_{c}(m,\Lambda)/\alpha_{c}(0)$ (solid line) and the normalized critical flavor number $N_{c}(m,\Lambda,g)/N_{c}(g)$ (dashed line) on the dimensionless ratio $m/\Lambda$. The calculation uses $g=10$, with $\alpha_{c}(0)= \alpha_c(0,\Lambda) =\pi/8$ and $N_{c}(g)$ (given by Eq.~\eqref{eq:Nc_PQED_limit}) for normalization, representing critical parameters of PQED. The monotonic increase of $\alpha_{c}$ and the corresponding decrease of $N_{c}$ with $m/\Lambda$ illustrate the screening-induced suppression of dynamical mass generation.}
    \label{fig:critical_behavior}
\end{figure}

In Fig.~\ref{fig:critical_behavior}, for ratios $m/\Lambda$ approaching $0.30$, the critical flavor number $N_c(m, \Lambda, g)$ decreases by approximately $10\%$ relative to the massless PQED limit, indicating that Yukawa screening tends to suppress dynamical mass generation. This behavior is corroborated by the corresponding increase in the critical coupling constant $\alpha_c(m, \Lambda)$ obtained in the quenched approximation, which still allows for chiral symmetry breaking.

The rainbow-unquenched approximation thus confirms that the pseudo-Proca screening scale controls the long-range structure of the interaction and shifts the location of the chiral critical point. In the next section, we go beyond the bare rainbow truncation and examine how this qualitative picture is modified when the fermion--gauge vertex is improved through the Ball--Chiu construction.

\section{Beyond the Rainbow Approximation: The Ball-Chiu Vertex}
\label{sec:BallChiu}

Building on the results obtained in Secs.~\ref{sec:RQApprox} and~\ref{sec:RUnQApprox}, we now examine whether the suppression of dynamical mass generation by the pseudo-Proca screening scale remains stable once vertex corrections consistent with the Ward--Takahashi identity are included. To this end, we adopt the Ball--Chiu construction, which provides a minimal nonperturbative extension of the fermion--gauge-boson vertex while preserving the gauge-covariant structure of the Schwinger--Dyson equations. Our aim here is not to derive fully closed analytical expressions, but rather to assess the robustness of the critical behavior found in the previous sections when wavefunction dressing effects are taken into account.

A well-established ansatz that satisfies the Ward-Takahashi identity and which is  free of kinematic singularities is the Ball-Chiu vertex~\cite{roberts1994dyson,BallChiu1980}
\begin{align}
\Gamma_{\mu}^{BC}(p,k) &= - f_{0}\gamma_{\mu} - f_{1}(\slashed{p}+\slashed{k})(p+k)_{\mu} + f_{2}(p+k)_{\mu}, \label{eq:BC_vertex}
\end{align}
where the scalar functions $f_0$, $f_1$, and $f_2$ are given by
\begin{align}
f_{0} &= \frac{A(p)+A(k)}{2}, \nonumber \\[0.2cm]
f_{1} &= \frac{A(p)-A(k)}{2(p^{2}-k^{2})}, \nonumber \\[0.2cm]
f_{2} &= \frac{\Sigma(p)-\Sigma(k)}{p^{2}-k^{2}}.
\label{eq:BC_functions}
\end{align}

In contrast to the standard rainbow approximation, which assumes $A(p) \approx 1$, the inclusion of the Ball-Chiu vertex yields a non-linear, coupled system of integral equations for both $\Sigma(p)$ and $A(p)$. For PPQED, using the \textit{quenched} framework $\Delta_{\mu \nu}(p) \to \Delta_{\mu \nu}^{(0)}(p)$ in the Landau gauge ($\lambda \to \infty$) and the Ball-Chiu vertex, inserting Eq.~\eqref{eq:BC_vertex} in Eqs.~\eqref{eq:sigma_initial} and~\eqref{eq:A_initial}, the gap equations are given by
\begin{align}
    \Sigma(p) &= \frac{\alpha}{\pi} \int_0^{\Lambda} \frac{dk \, k^2}{A^2(k) k^2 + \Sigma^2(k)} \nonumber \\
    &\quad \times \Big[ f_0 \Sigma(k) I_1 + p^2 k^2 \big(2 \Sigma(k) f_1 - A(k) f_2\big) I_2 \Big], \label{eq:BC_Sigma}\\
    A(p) &= 1 + \frac{\alpha}{\pi p^2} \int_0^{\Lambda} \frac{dk \, k^2}{A^2(k) k^2 + \Sigma^2(k)} \nonumber \\
    &\quad \times \Big[ A(k) f_0 p k I_3 - p^2 k^2 I_2 \nonumber \\
    &\quad \times \big( A(k) f_0 + A(k) f_1 (p^2 + k^2) + \Sigma(k) f_2 \big) \Big], \label{eq:BC_A}
\end{align}
where $I_1$, $I_2$, and $I_3$ are the angular integrals 
\begin{align}
    I_1 &= \int_{0}^{\pi}
\frac{\sin\theta}
{\sqrt{q^2(\theta)+m^{2}}}
\,d\theta, \\[0.2cm]
   I_2 &=  \int_{0}^{\pi}
\frac{\sin\theta\,\bigl(1-\cos^{2}\theta\bigr)}
{q^2(\theta)\sqrt{q^2(\theta)+m^{2}}}
\,d\theta, \\[0.2cm]
    I_3 &= \int_{0}^{\pi}
\frac{\sin\theta\,\cos\theta}
{\sqrt{q^2(\theta)+m^{2}}}
\,d\theta,
\end{align}
and $q^2(\theta) = p^2+k^2- 2 p k \cos{\theta}$.

\begin{figure}[htpb]
    \centering
    \begin{minipage}{0.48\linewidth}
        \centering
        \includegraphics[width=\linewidth]{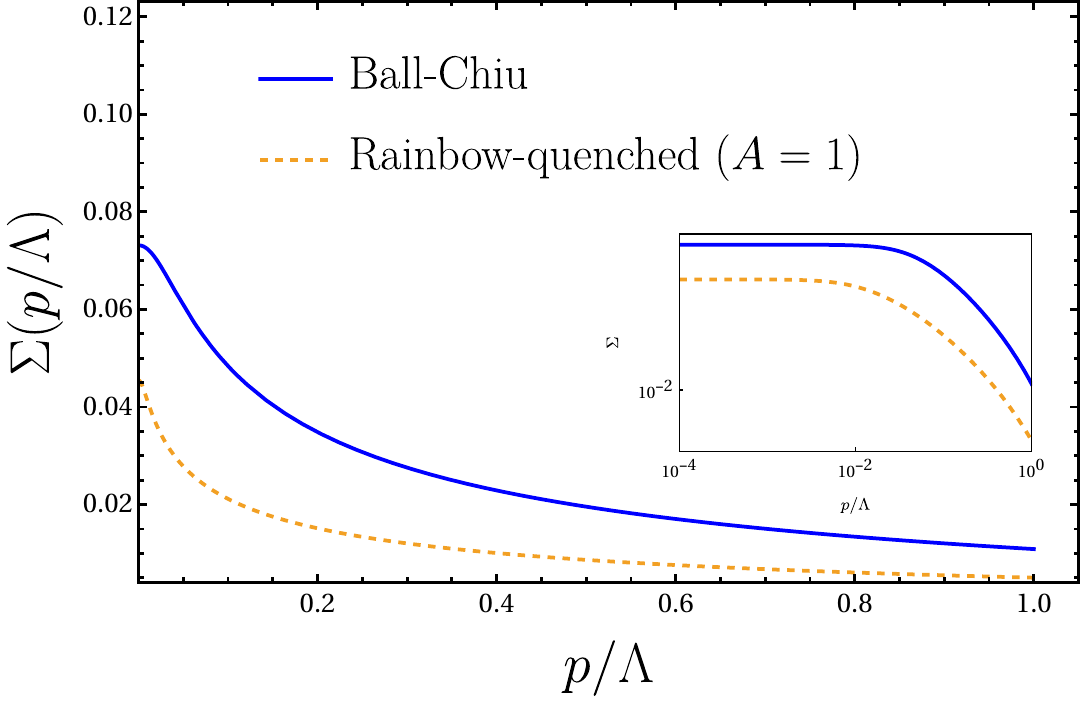}
        \vspace{0.1cm}
        
        \textbf{(a)}
        \label{fig:bc_q_sigma}
    \end{minipage}\hfill
    \begin{minipage}{0.48\linewidth}
        \centering
        \includegraphics[width=\linewidth]{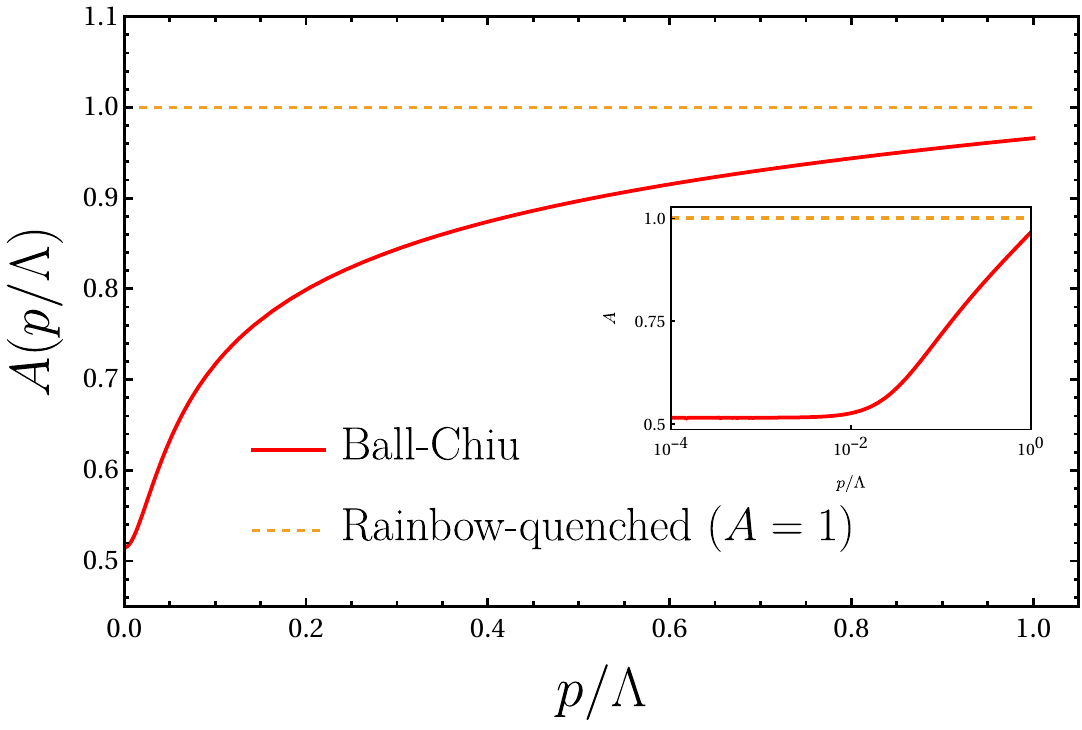}
        \vspace{0.1cm}
        
        \textbf{(b)}
        \label{fig:bc_q_a}
    \end{minipage}
    
    \caption{Comparison between the Ball-Chiu and rainbow-quenched approximations in PPQED, evaluated at $m = 1 \times 10^{-5}\, \Lambda$ and $\alpha = 0.39$. (a) The mass function $\Sigma(p/\Lambda)$ shows a quantitative enhancement in the deep IR regime when vertex corrections are included. (b) The wavefunction renormalization $A(p/\Lambda)$ is heavily suppressed at low momenta, deviating from the bare approximation $A(p)=1$. The insets display the corresponding quantities with a logarithmic momentum scale to emphasize the asymptotic behavior in the deep infrared and ultraviolet limits.}
    \label{fig:BC_quenched}
\end{figure}

\begin{figure}[htpb]
    \centering
    \begin{minipage}{0.48\linewidth}
        \centering
        \includegraphics[width=\linewidth]{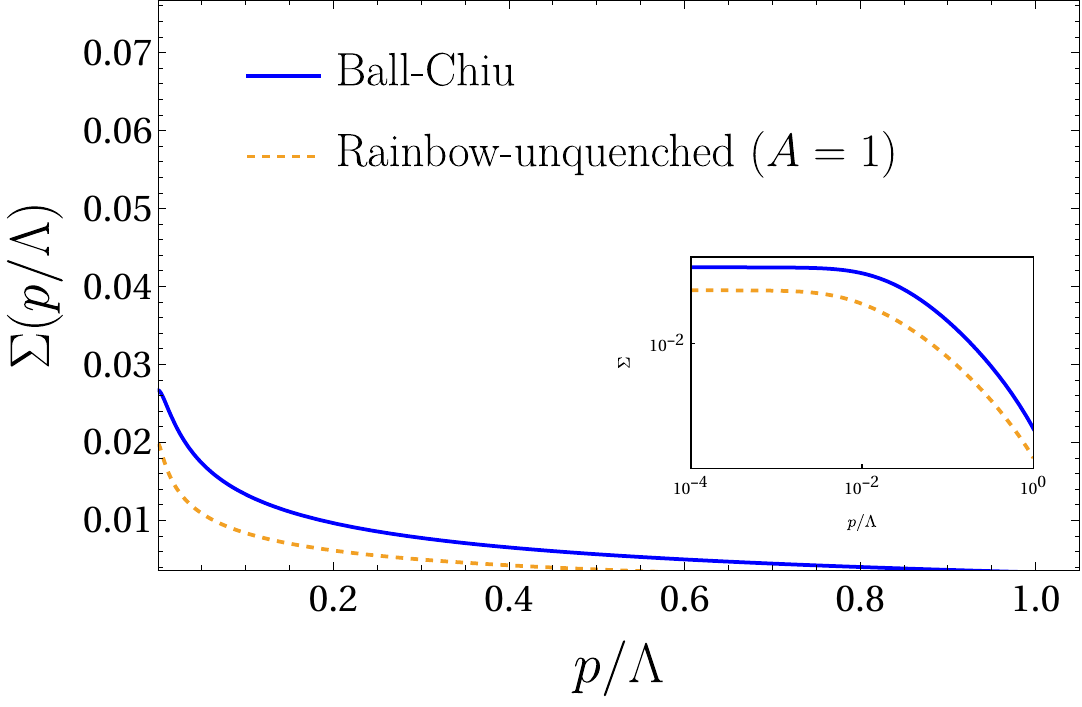}
        \vspace{0.1cm}
        
        \textbf{(a)}
        \label{fig:bc_u_sigma}
    \end{minipage}\hfill
    \begin{minipage}{0.48\linewidth}
        \centering
        \includegraphics[width=\linewidth]{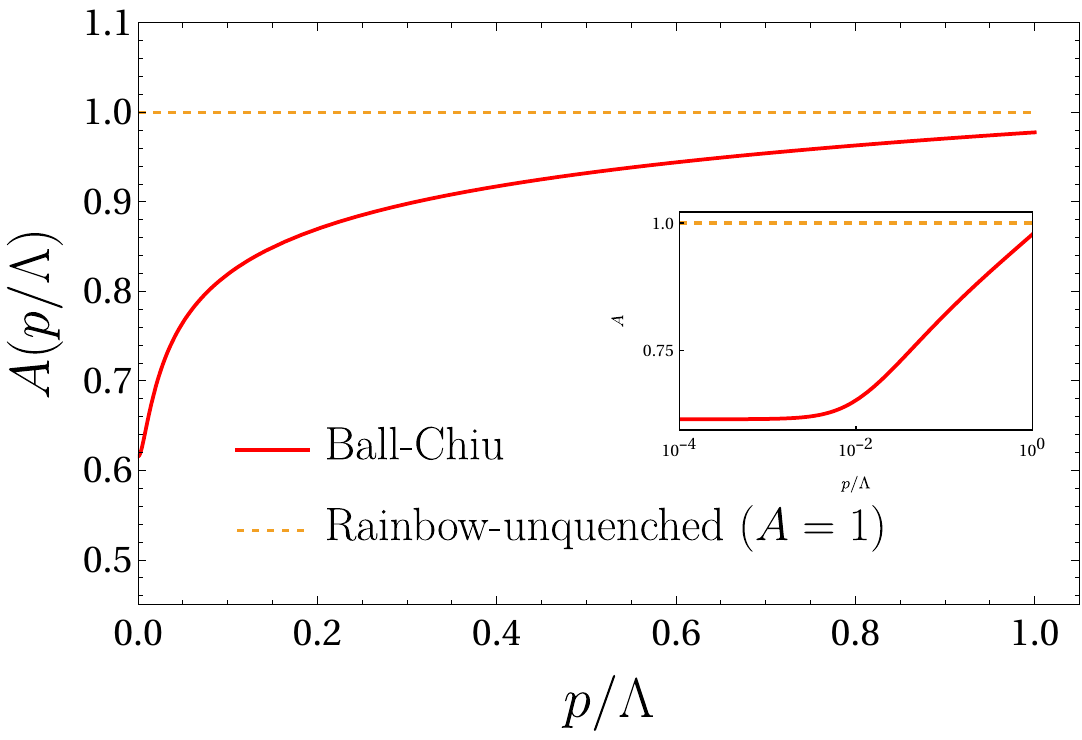}
        \vspace{0.1cm}
        
        \textbf{(b)}
        \label{fig:bc_u_a}
    \end{minipage}
    
    \caption{Comparison between the Ball-Chiu and rainbow-unquenched approximations, evaluated at $m = 1 \times 10^{-5}\, \Lambda$, $g = 100$, and $N = 4$. Similar to the quenched case, (a) the dynamically generated mass is enhanced in the IR limit, driven by (b) the suppression of the wavefunction renormalization $A(p/\Lambda) < 1$ at low momenta. The insets highlight the solutions across the entire momentum range using a logarithmic scale for the horizontal axis.}
    \label{fig:BC_unquenched}
\end{figure}

Extending this procedure to the \textit{unquenched} scenario (incorporating vacuum polarization via the $1/N$ expansion) preserves the structure of Eqs.~\eqref{eq:BC_Sigma} and~\eqref{eq:BC_A}. The only modifications are the replacement of $\alpha/\pi$ to the constant $g / (4 \pi^2 N)$, and the increasing of the internal screening kernel by the vacuum polarization, which modifies the denominator $\sqrt{q^2(\theta) + m^{2}} \to  \sqrt{q^2(\theta) + m^{2}} + \frac{g}{16} q(\theta) $.  The coupled integral equations were solved iteratively on a logarithmic momentum grid using a damped fixed-point scheme (with an under-relaxation factor of $\omega = 0.3$) to ensure stable convergence across the parameter space. Comprehensive details regarding the numerical regularization of the vertex singularities are reserved for~\ref{B}.  To isolate the effects of the non-perturbative vertex dressing, we compared the Ball-Chiu solutions against their respective ladder approximations. The numerical evaluations considered a fixed screening scale of $m = 1 \times 10^{-5}\, \Lambda$. We set $\alpha = 0.39$ for the quenched scenario, and $g = 100$ with $N = 4$ for the unquenched scenario. The numerical results obtained for the quenched and unquenched regimes are displayed in Fig.~\ref{fig:BC_quenched} and Fig.~\ref{fig:BC_unquenched}, respectively.

Introducing the Ball-Chiu vertex quantitatively modifies the infrared behavior of the system. As shown in Figs.~\ref{fig:BC_quenched}(a) and \ref{fig:BC_unquenched}(a), our numerical solutions indicate an enhancement of the dynamically generated mass gap at $p\to 0$ with respect to the corresponding rainbow approximations. Concurrently, as seen in Figs.~\ref{fig:BC_quenched}(b) and \ref{fig:BC_unquenched}(b), we observe a suppression of the wavefunction renormalization, with $A(p)$ decreasing below unity in the infrared regime. These numerical outcomes demonstrate that the bare assumption $A(p)\approx 1$ leads to an underestimate of the dynamically generated mass in the deep infrared region. Nevertheless, the main qualitative conclusion of the present work remains unchanged. Indeed, the scale $m$ still induces a Yukawa-like screening mechanism that suppresses chiral symmetry breaking, while the critical thresholds retain the same overall dependence on the screening scale $m$. Therefore, the vertex-improved analysis confirms that the scaling behavior obtained within the ladder truncations reflects a genuine feature of PPQED rather than an artifact of the simplest approximation.

Although the Ball--Chiu vertex induces quantitative corrections to the infrared behavior of the fermion propagator, it preserves the central qualitative conclusion of the manuscript: the pseudo-Proca screening parameter remains the quantity that controls the suppression of chiral symmetry breaking in PPQED. This result reinforces the interpretation advanced in the rainbow analysis. It is worth mentioning that the inclusion of this improved vertex interaction also could be realized for PQED (in the case $m=0$), where similar conclusions should be obtained. Next, we discuss an anisotropic extension of PPQED.

\section{Anisotropic extension}
\label{sec:Anisotropy}

The purpose of this section is not to provide a complete description of realistic Dirac materials. Rather, it serves as a consistency test of how the pseudo-Proca screening mechanism behaves when Lorentz symmetry is explicitly broken in the fermionic sector. In such systems, Lorentz invariance is explicitly broken in the matter sector, and the chiral dynamics depends not only on the pseudo-Proca screening parameter $m$, but also on the ratio $v_F/c$. To incorporate this effect, we generalize the fermionic sector by replacing the relativistic Dirac operator according to $i\gamma^\mu\partial_\mu \longrightarrow i\gamma^0\partial_0 + iv_F\gamma^i\partial_i$, with $i=1,2$. To keep track of the non-relativistic hierarchy $v_F/c \ll 1$ and to make the scaling relations dimensionally transparent, we temporarily restore explicit factors of $c$ in the anisotropic sector.

In Euclidean space, the corresponding bare fermion propagator becomes
\begin{equation}
S_{F}^{(0)}(p)
=
\frac{
\gamma^{0}p_{0}
+
v_{F}\gamma^{i}p_{i}
+
Mc^{2}
}{
p_{0}^{2}
+
v_{F}^{2}\mathbf{p}^{2}
+
M^{2}c^{4}
},
\label{eq:SF0_aniso}
\end{equation}

The bare pseudo-Proca propagator retains the same non-local structure as in Sec.~\ref{sec:Model}, but introducing the appropriate dimensions, we find
\begin{equation}
\Delta_{\mu\nu}^{(0)}(p)
=
\frac{c}{2\sqrt{p_0^{2}+\mathbf{p}^2 c^2 +m^{2}c^{4}}}
\left(
\delta_{\mu\nu}
-
\frac{p_{\mu}p_{\nu}}{p^{2}}
\right).
\label{eq:D0_aniso}
\end{equation}
The explicit factor of $c$ in the gauge-field propagator restores dimensional consistency, which is an essential requirement in $(2+1)$D theories without full Lorentz invariance. Similar investigations have recently been performed within the framework of anisotropic QED$_4$~\cite{isobe2012theory,isobe2013renormalization,isobe2016coulomb}. The interaction factor at the fermion--gauge vertex becomes anisotropic and may be written as
\begin{equation}
e\Gamma_{(0)}^{\mu},
\qquad
\Gamma_{(0)}^{\mu}
= -
\left(  
\gamma^{0},
 \frac{v_F}{c}\gamma^{i}
\right).
\label{eq:vertex_aniso}
\end{equation}
This form reflects the asymmetry between temporal and spatial components and ensures that the interaction term has the appropriate units and symmetry properties.

Similarly to the isotropic case, our main object of interest is the fermion propagator. In the anisotropic regime, however, it is convenient to distinguish the renormalization of the spatial kinetic term from the mass function. We therefore parametrize the full inverse propagator as follows
\begin{equation}
S_{F}^{-1}(p)
=
-\gamma^{0}p_{0}
-
B(p)\,v_{F}\gamma^{i}p_{i}
+
\Sigma(p)c^{2},
\label{eq:SF_aniso}
\end{equation}
where $B(p)$ is the Fermi-velocity renormalization function and $\Sigma(p)$ is the mass function.

Applying tracing operations in Eq.~\eqref{eq:SDE_fermion} and~\eqref{eq:SF_aniso} leads us to the following results
\begin{align}
    \Sigma(p)  &=  M + \frac{e^2}{c^2 \Tr{\mathbb{I}}} \int \frac{d^3 k}{(2 \pi)^3} \nonumber \\
    & \qquad \times \Tr{ \gamma^{\mu} S_F(k) \Gamma^{\nu}(k,p)  } \Delta_{\mu \nu}(p-k) \label{eq:sigma_aniso_initial} \\[0.3cm]
    B(p) &= 1 + \frac{\pi \alpha}{ \mathbf{p}^2 } p_i  \int\frac{d^3k}{(2 \pi)^3} \nonumber \\
    &\qquad \times \Tr\{\gamma^i \gamma^{\mu} S_F(k) \Gamma^{\nu}(k,p)\} \Delta_{\mu \nu}(p-k). \label{eq:B_aniso_initial}
\end{align}

In analogy to the approximation adopted for $A(p)$ in the isotropic case, we set $B(p)\simeq 1$ throughout the analytical treatment. This assumption is supported numerically within the parameter range of interest (see~\ref{A}), but some important deviations are found in the regime $p\to0$, where Fermi velocity renormalization occurs.

Initially, we take $M=0$ and evaluate the fermion self-energy within the rainbow-quenched approximation and in the static regime. More precisely, we retain only the instantaneous part of the gauge interaction, so that $\Delta_{\mu\nu}^{(0)}(p)\rightarrow \Delta_{00}^{(0)}(p_0=0,\mathbf{p})$ and $\Gamma^\mu(k,p)\rightarrow \Gamma_{(0)}^\mu \rightarrow -\gamma^0$. This approximation is physically motivated by the anisotropic hierarchy $v_F\ll c$, which suppresses dynamical corrections of order $O(v_F/c)$. Accordingly, the present treatment is intended to capture the leading trend of the anisotropic critical coupling rather than to provide a fully dynamical description of the anisotropic gap equation. In this sense, the anisotropic analysis should be understood as a leading-order exploratory treatment, since retardation effects and self-consistent velocity dressing are not included at this stage. This static treatment is also consistent with well-established analyses of graphene and related two-dimensional materials, where the Coulomb interaction is predominantly instantaneous~\cite{Kovner1990,popovici2013fermi,Dorey1992,Gonzales1994}.

Applying these assumptions in Eq.~\eqref{eq:sigma_aniso_initial}, the mass function reduces to
\begin{equation}
\Sigma(p)
=
\pi v_F \alpha
\left(
1+\frac{v_F^2}{c^2}
\right)
\int_{0}^{\Lambda}
\frac{dk}{(2\pi)^2}
\,
\frac{k\,\Sigma(k)\,\mathcal{M}_{A}(k,p)}
{\sqrt{v_F^{2}k^{2}+\Sigma^{2}(k)c^{4}}},
\label{eq:gap_aniso_integral}
\end{equation}
where the angular kernel is
\begin{equation}
\mathcal{M}_{A}(k,p)
=
\int_{0}^{2\pi}
\frac{d\theta}
{\sqrt{p^{2}+k^{2}-2pk\cos\theta+m^{2}c^{2}}}.
\label{eq:MA_kernel}
\end{equation}
As in the isotropic case, the pseudo-Proca screening scale appears in the denominator of the kernel and, therefore, suppresses the interaction responsible for dynamical mass generation. The angular integral in Eq.~\eqref{eq:MA_kernel} can be expressed in terms of complete elliptic integrals of the first kind~\cite{NIST2010,gradshteyn2014table}, although its explicit closed form is not needed for the discussion of the asymptotic scaling behavior developed below.

To extract the critical structure of the anisotropic gap equation, we proceed as in Sec.~\ref{sec:RQApprox} and consider the regime in which the mass function is small compared to the kinetic term, $\Sigma^{2}(p)\ll \left(\frac{v_F}{c^{2}}\right)^{2}p^{2}$.

After differentiating the integral equation with respect to $p$, one arrives at the differential form
\begin{align}
\frac{d}{dp}
\left[
\frac{(p^{2}+m^{2})^{3/2}}{p}\,
\Sigma'(p)
\right] + \frac{  \alpha}{2}   \left(   1  + \frac{v_F^2}{c^2}    \right)  \Sigma(p)=0.
\label{eq:gap_aniso_diff}
\end{align}
This equation should be interpreted in the same way as its isotropic counterpart: its role is to expose the scaling structure of the problem and to provide an analytical estimate for the anisotropic critical coupling.

In the ultraviolet-dominated regime, and assuming once more the hierarchy $m\ll \Lambda$, we approximate
\begin{equation}
p^{2}
\left(
1+\frac{m^{2}}{p^{2}}
\right)^{3/2}
\approx
p^{2}
\left(
1+\frac{m^{2}}{\Lambda^{2}}
\right)^{3/2}.
\label{eq:aniso_UV_approx}
\end{equation}
Under this approximation, Eq.~\eqref{eq:gap_aniso_diff} has the power-law solution

\begin{equation}
\Sigma(p)
=
\bar{C}_{1}\,p^{-\bar{\lambda}_{1}/2}
+
\bar{C}_{2}\,p^{\bar{\lambda}_{1}/2-1},
\label{eq:Sigma_aniso_solution}
\end{equation}
where
\begin{equation}
\bar{\lambda}_{1}
=
1-\sqrt{1-\frac{\alpha}{\alpha_{c}^{*}}}.
\label{eq:lambdabar1}
\end{equation}
The anisotropic critical coupling is therefore given by
\begin{equation}
\alpha_c^{*} = \frac{1}{2}  \frac{\left( 1+ \frac{m^2}{\Lambda^2}\right)^{3/2}}{ \left(   1  + \frac{v_F^2}{c^2}    \right)}.
\label{eq:alpha_c_aniso}
\end{equation}

Eq.~\eqref{eq:alpha_c_aniso} shows that, within the present static approximation, the anisotropic critical coupling scales inversely with the factor $(1+v_F^2/c^2)$. Therefore, for fixed $m$ and $\Lambda$, increasing $v_F$ tends to reduce $\alpha_c^*$. Nevertheless, for the physically relevant range of two-dimensional materials, namely $v_F/c\sim 10^{-3}$--$10^{-2}$~\cite{Katsnelson2007,Wang2019}, one still finds $\alpha_c^*>\alpha_c$ when Eq.~\eqref{eq:alpha_c_aniso} is compared with the isotropic result in Eq.~\eqref{eq:alpha_c_quenched}. In this sense, the static anisotropic approximation yields a symmetry-breaking threshold that remains above the isotropic one in the regime of interest.

\section{Summary and outlook}
\label{sec:Conclusion}

In this work, we investigated dynamical mass generation in PPQED by means of Schwinger--Dyson equations in quenched, unquenched, and vertex-improved truncation schemes. Our results show that the pseudo-Proca  screening parameter $m$ acts as a continuous screening scale in the reduced $(2+1)$D theory: as $m$ increases, the interaction becomes more Yukawa-like, the effective kernel is weakened, and chiral symmetry breaking is suppressed. In the rainbow-quenched approximation, this appears as an increase in the critical coupling $\alpha_c(m,\Lambda)$, whereas in the rainbow-unquenched case it is reflected in the corresponding dependence of the critical flavor number $N_c(m,\Lambda,g)$.

The unquenched analysis and the Ball–Chiu construction further suggest that the qualitative trends found in the simplest truncation remain stable when basic vacuum-polarization and vertex corrections are incorporated. Although vertex and vacuum-polarization effects modify the quantitative values of the critical parameters, the qualitative conclusion remains unchanged: the pseudo-Proca scale monotonically suppresses dynamical mass generation. In this sense, the rainbow results provide analytically transparent baseline estimates, while the Ball--Chiu solutions offer a more realistic non-perturbative point of comparison. To the best of our knowledge, analogous vertex-improved studies remain relatively unexplored in reduced gauge theories such as PQED or PPQED. Its correction, however, seems to be similar to those of  QED$_4$ and QED$_3$ discussed in earlier studies~\cite{Bashir1994,Maris1996}.

We also considered an anisotropic extension motivated by planar Dirac materials. Within the static approximation adopted in Sec.~\ref{sec:Anisotropy}, the corresponding critical coupling scales as $\alpha_c^*\propto \alpha_c/(1+v_F^2/c^2)$, showing that the critical behavior depends not only on the pseudo-Proca screening scale but also on the hierarchy between the fermionic and gauge-sector velocities. This part of the analysis should be viewed as exploratory, but it captures the leading effect of anisotropy on the critical structure of PPQED.

In general, our results highlight the role of the screening parameter $m$ in modulating the non-perturbative critical behavior of PPQED.
Future work should incorporate more complete vertex and propagator dressings, finite-temperature and finite-density effects, and a more refined treatment of anisotropy. A possible conceptual connection with systems exhibiting Proca-like interaction may provide useful qualitative intuition~\cite{Said2021} and with the introduction of new parameters of the material we are likely to obtain more quantitative results. This will be the subject of future investigation.

\appendix

\section{Numerical Solutions for the Rainbow Vertex}
\label{A}

This Appendix summarizes the numerical procedure used to solve the integral gap equations discussed in Secs.~\ref{sec:RQApprox} and~\ref{sec:RUnQApprox}, as well as the consistency checks associated with the renormalization functions. The non-linear equations for the mass function $\Sigma(p)$ and for the dressing functions $A(p)$ and $B(p)$ (given by Eqs.~\eqref{eq:A_initial} and~\eqref{eq:B_aniso_initial}, respectively) in the rainbow-quenched regime were discretized by means of repeated trapezoidal quadrature~\cite{Atkinson_1997}, which is well suited to the smooth kernels considered here.

To properly capture both the deep infrared (IR) and ultraviolet (UV) behaviors, the momentum grid was established with $N_h = 300$ intervals logarithmically spaced over $10^{-3} \leq p/\Lambda \leq 1$. In our numerical evaluations, this sets the IR cutoff to $p_0 = 10^{-3}\Lambda$ and the UV cutoff to $\Lambda = 10$.

\begin{figure}[htpb]
    \centering
    \includegraphics[width=0.85\linewidth]{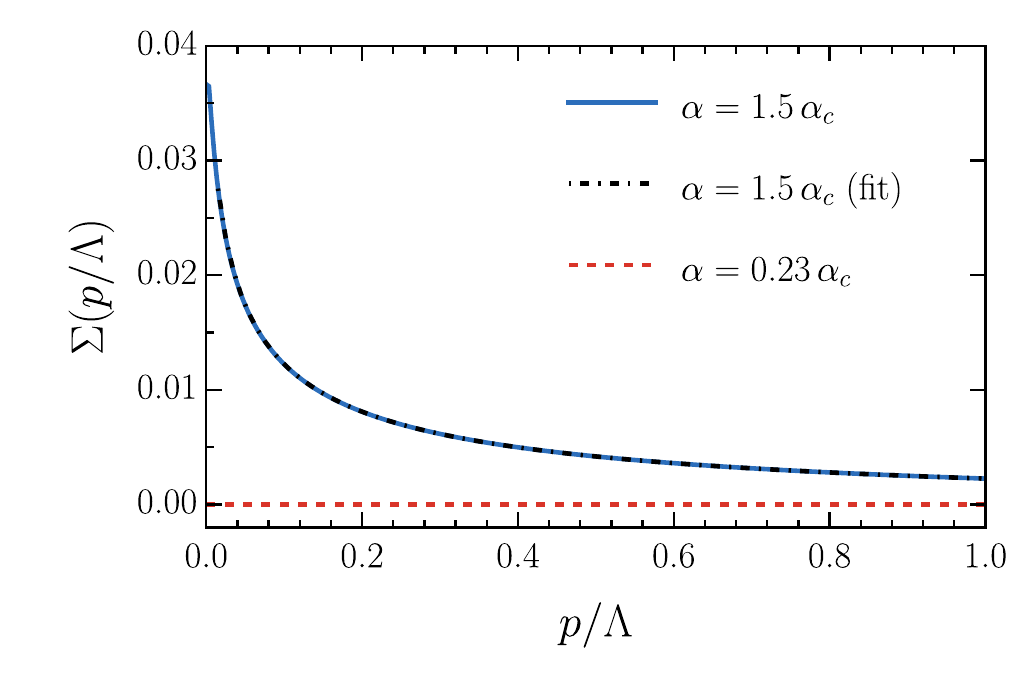}
    \textbf{(a)} $m = 1 \times 10^{-5}\, \Lambda$
    
    \vspace{0.4cm}
    
    \includegraphics[width=0.85\linewidth]{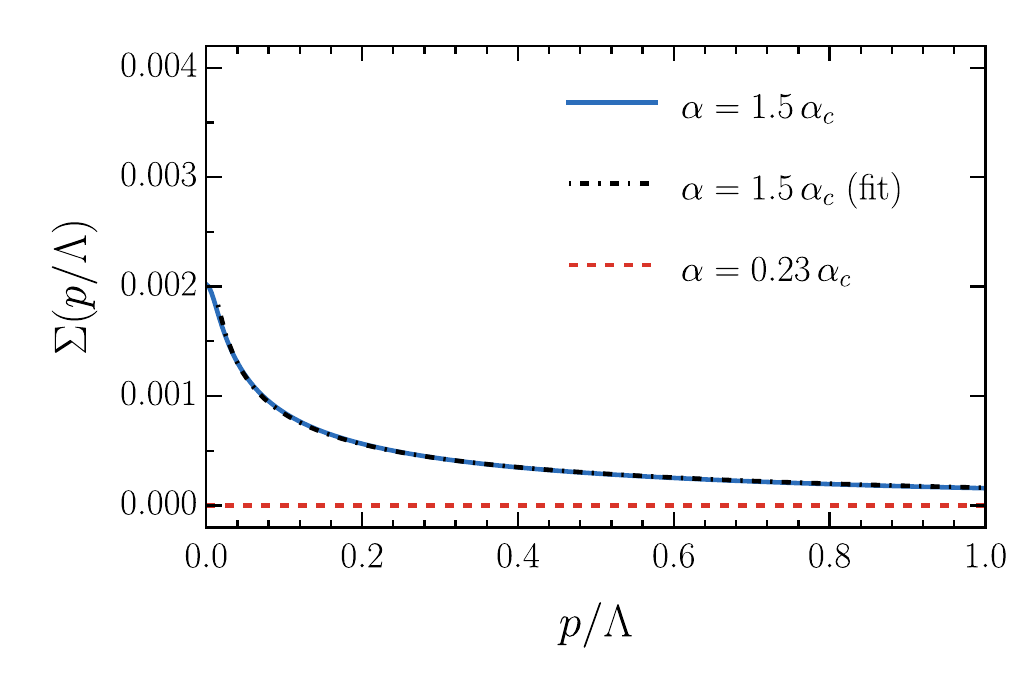}
    \textbf{(b)} $m = 1 \times 10^{-2}\, \Lambda$
    
    \caption{Numerical solution of the mass function $\Sigma(p)$ from Eq.~\eqref{eq:gap_quenched_2} in the rainbow-quenched approximation. Panels (a) and (b) correspond to different screening scales $m$. For each value, two values of the coupling constant $\alpha$ are shown: one below and other above the analytical critical coupling $\alpha_c \approx 0.39$. The black curve represents the analytical solution from the linearized differential equation [Eq.~\eqref{eq:gap_power_solution}], with parameters $A_1$ and $A_2$ fitted at $\alpha = 1.5 \ \alpha_c$.}
    \label{fig:graficosquenchedsigma}
\end{figure}

In the rainbow-quenched approximation, our numerical results accurately capture the chiral phase transition. As the screening scale vanishes ($m \to 0$), the critical coupling correctly recovers the analytical pseudo-QED (PQED) limit, $\alpha_c(m \to 0, \Lambda) = \alpha_c^{\text{PQED}} = \pi/8$~\cite{alves2013chiral}. As expected from the integral kernel, the dynamically generated mass reaches its maximum at the IR cutoff, $\Sigma(p_0) \geq \Sigma(p)$, and rapidly vanishes towards the UV limit, $\Sigma(\Lambda) \ll \Sigma(p_0)$. Fig.~\ref{fig:graficosquenchedsigma} illustrates this critical behavior for different screening scales $m$, matching the analytical curves whose integration constants were determined via least-squares fitting to the numerical data [see Eq.~\eqref{eq:gap_power_solution}]. From these results, we can conclude that the mass function decreases when either we increase the screening scale or decrease the interaction.

\begin{figure}[htpb]
    \centering
    \includegraphics[width=0.85\linewidth]{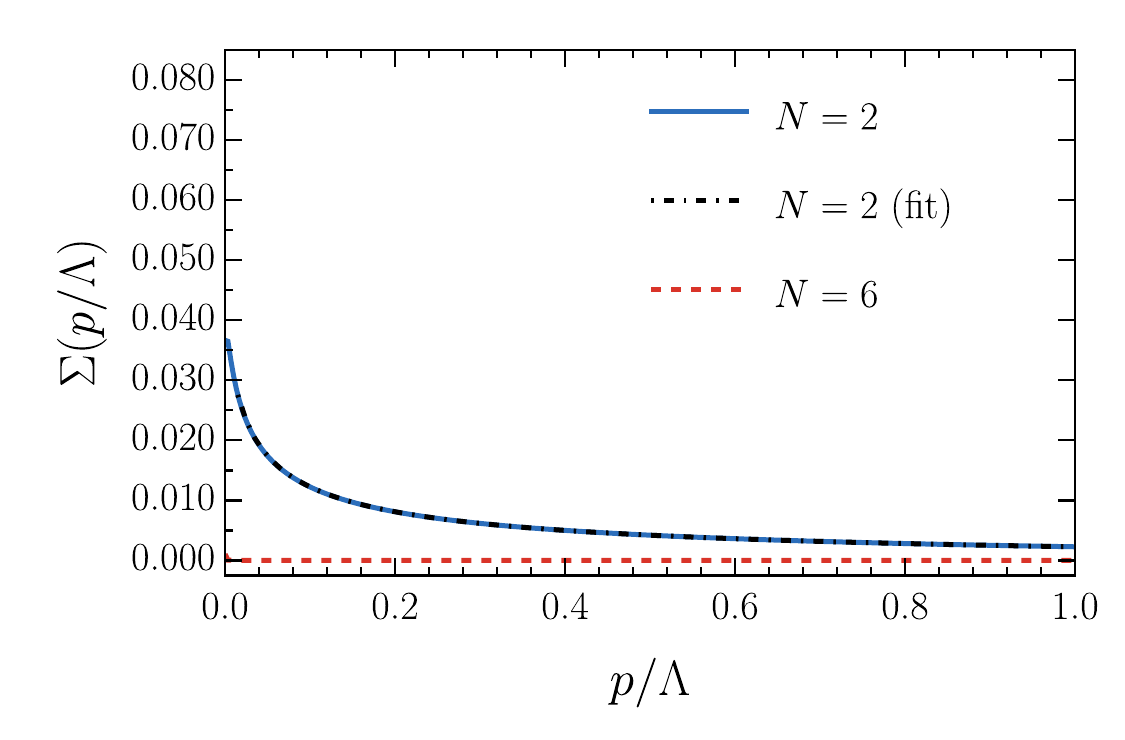}
    \textbf{(a)} $m = 1 \times 10^{-5}\, \Lambda$
    
    \vspace{0.4cm}
    
    \includegraphics[width=0.85\linewidth]{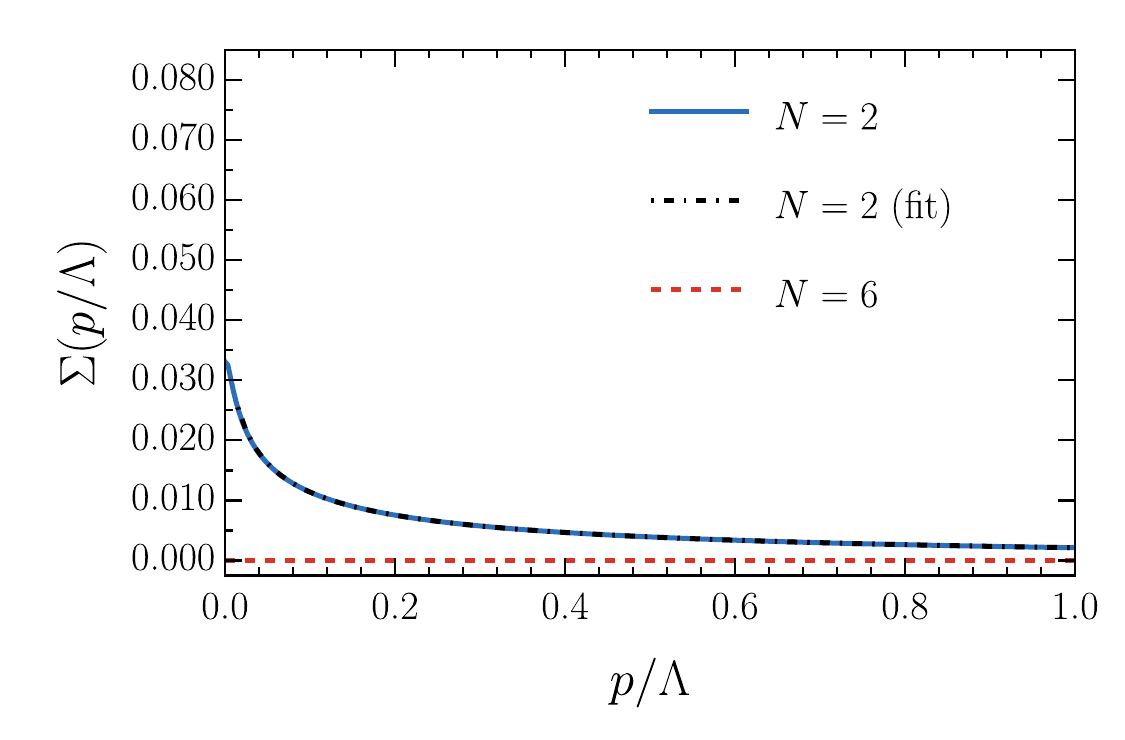}
    \textbf{(b)} $m = 1 \times 10^{-2}\, \Lambda$
    
    \caption{Numerical solution of the mass function $\Sigma(p)$ from Eq.~\eqref{eq:gap_unquenched_integral} in the rainbow-unquenched approximation, with a fixed critical flavor number $N_c = 3$ and $g=198$. The blue and red curves show results for $N = 2$ and $N = 6$, illustrating dynamical mass generation below $N_c$ and symmetry restoration above $N_c$, respectively. The black curve corresponds to the analytical solution from the linearized differential equation [Eq.~\eqref{eq:Sigma_power_unquenched}], with parameters $C_1$ and $C_2$ fitted for $N = 2$.}
    \label{fig:graficosunquenchedsigma}
\end{figure}

For the rainbow-unquenched scenario, the mass function $\Sigma(p)$ exhibits a similar continuous phase transition, but this time driven by the critical flavor number $N_c$, as depicted in Fig.~\ref{fig:graficosunquenchedsigma}. The analytical solutions from Eq.~\eqref{eq:Sigma_power_unquenched} are again in excellent agreement with the numerical data. Here, we conclude that the mass function decreases as we increase the flavor number of fermions.

\begin{figure}[htpb]
    \centering
    \includegraphics[width=0.85\linewidth]{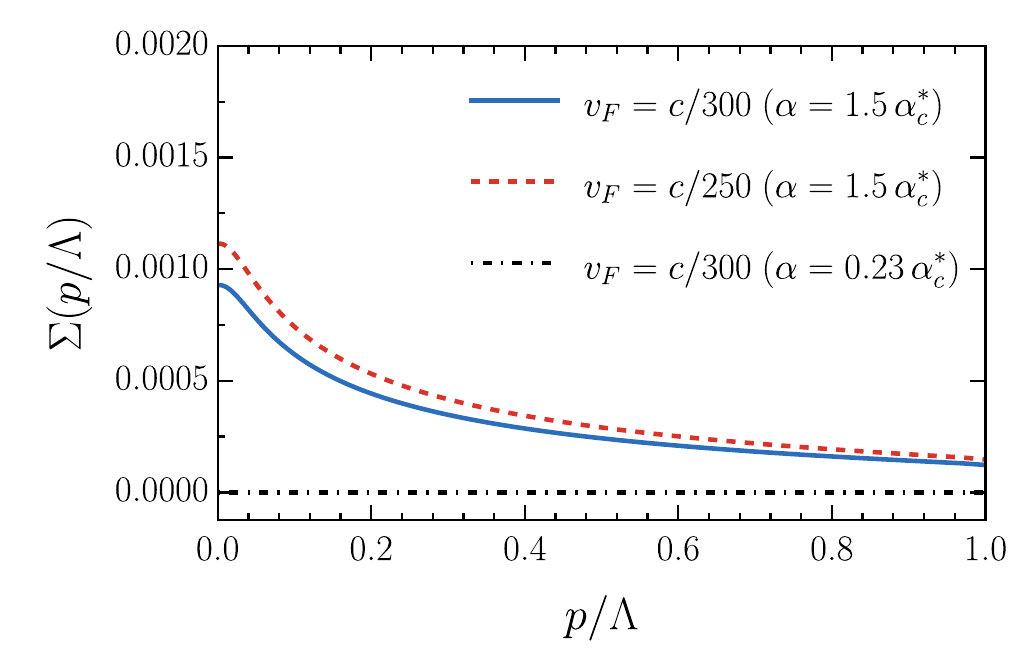}
    \textbf{(a)} $m = 1 \times 10^{-5}\, \Lambda$
    
    \vspace{0.4cm}
    
    \includegraphics[width=0.85\linewidth]{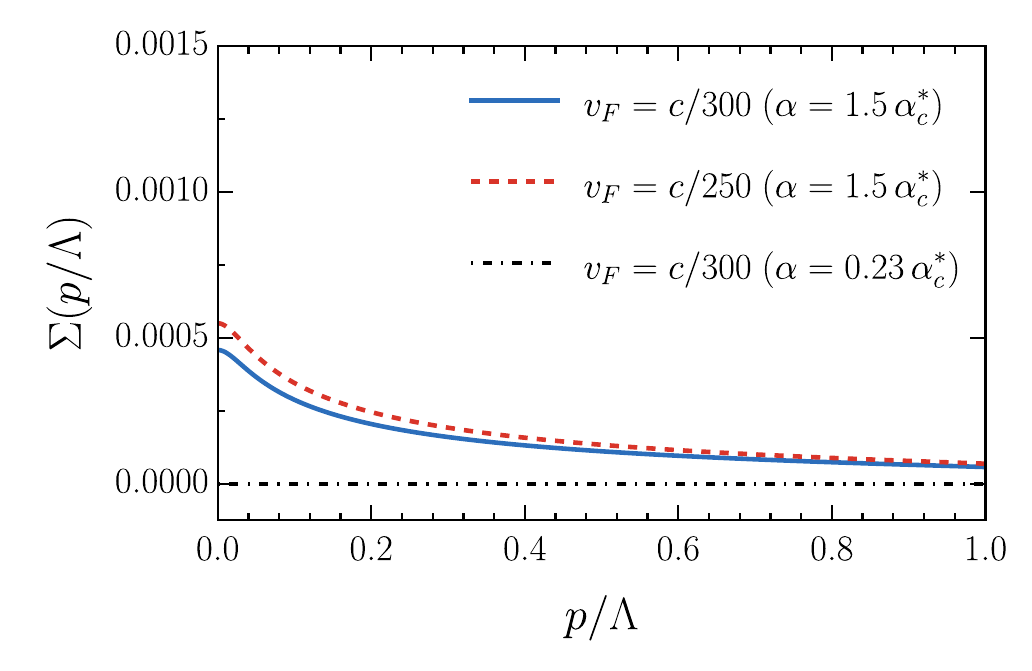}
    \textbf{(b)} $m = 1 \times 10^{-1}\, \Lambda$
    
    \caption{Numerical solutions of Eq.~\eqref{eq:gap_aniso_integral} are presented for two screening scale values and two Fermi velocities ($v_F = c/300$ and $v_F = c/250$). Each plot includes curves for two coupling constants: $\alpha = 0.23 \ \alpha_c^*$ (black) and $\alpha = 1.5 \ \alpha_c^*$ (blue and red), facilitating the analysis of critical behavior within the model.}
    \label{fig:anisograficosquenchedsigma}
\end{figure}

The same numerical framework can be extended to the anisotropic case, in which the Fermi velocity $v_F \neq c$ modifies the mass function (see Eq.~\eqref{eq:gap_aniso_integral}). Fig.~\ref{fig:anisograficosquenchedsigma} illustrates the corresponding mass generation for different screening scale regimes and Fermi velocities, showing the associated shift of the critical coupling threshold. Indeed, we have shown in Eq.~\eqref{eq:alpha_c_aniso} that in this regime the critical coupling constant is very close to $0.5$, which is larger than the isotropic value of this quantity.

\begin{figure}[htpb]
    \centering
    \includegraphics[width=0.85\linewidth]{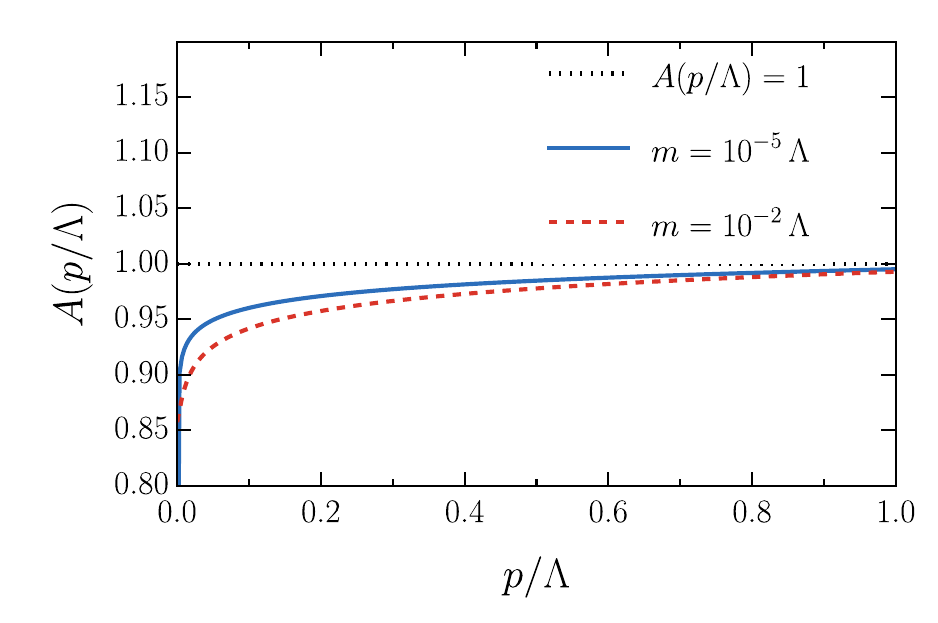}
    \caption{Numerical solutions for the wavefunction renormalization $A(p)$ in the symmetric phase for different screening scale values. Evaluated at $\alpha = 0.1$.}
    \label{fig:graficosAquenched}
    
    \vspace{0.6cm}
    
    \includegraphics[width=0.85\linewidth]{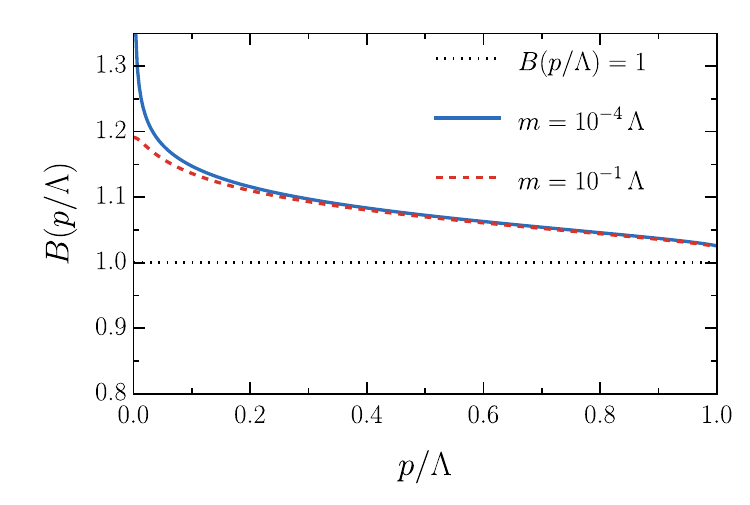}
    \caption{Numerical solutions for the Fermi velocity renormalization $B(p)$ in the symmetric phase. Evaluated at $\alpha = 0.1$ and $v_F = c/300$.}
    \label{fig:graficosAnisoWFRquenched}
\end{figure}

Finally, in order to test the consistency of the bare-vertex assumptions in the symmetric phase, $\Sigma(p)=0$, we directly solved the corresponding integral equations (using the rainbow-quenched approximation) for the renormalization functions, given by Eq.~\eqref{eq:A_initial} for the isotropic case and~\eqref{eq:B_aniso_initial} for the anisotropic case. Fig.~\ref{fig:graficosAquenched} shows that, in the isotropic regime and for subcritical couplings such as $\alpha=0.1$, one finds $A(p)\approx 1$ throughout the momentum range considered. Likewise, Fig.~\ref{fig:graficosAnisoWFRquenched} indicates that the Fermi-velocity renormalization remains close to unity, $B(p)\approx 1$, over the entire interval analyzed. These results support the approximations employed in the analytical treatment.

\section{Numerical Evaluation of the Coupled SDEs with the Ball-Chiu Vertex}
\label{B}

The inclusion of the non-perturbative Ball-Chiu vertex in the Schwinger-Dyson equations leads to a highly non-linear coupled system for the mass function $\Sigma(p)$ and the wavefunction renormalization $A(p)$. In contrast to the rainbow treatment described in the~\ref{A}, the present system requires a different numerical strategy because of the kinematic singularities introduced by the vertex structure and the need for off-grid evaluations of the dressed functions.

In order to obtain $\Sigma(p)$ and $A(p)$, the external and internal momenta were discretized on a logarithmic grid of $N_h = 300$ points, spanning from a deep infrared cutoff $p_{0} = 10^{-3} \, \Lambda$ up to the ultraviolet cutoff $\Lambda = 10$. The logarithmic scale is crucial to properly capture the infrared dynamics where dynamical mass generation occurs. 

At each iteration step, the functions $\Sigma(k)$ and $A(k)$ were reconstructed over the full momentum domain by means of natural cubic splines. This interpolation is required both for the evaluation of the shifted momentum variable $q(\theta)=\sqrt{p^2+k^2-2pk\cos\theta}$ entering the angular integrals and for the regular treatment of the apparent kinematic singularities associated with the Ball-Chiu functions $f_1(p,k)$ and $f_2(p,k)$ in the limit $k\to p$. In order to avoid numerical instabilities and spurious divisions by zero during the adaptive integration, we replaced the corresponding finite-difference expressions by their L'Hôpital limits whenever $|p-k|/\Lambda<10^{-5}$, with the derivatives computed from the spline interpolation:

\begin{align}
    \lim_{k \to p} f_1(p,k) &= \frac{1}{4p} \frac{d A(p)}{dp}, \\
    \lim_{k \to p} f_2(p,k) &= \frac{1}{2p} \frac{d \Sigma(p)}{dp}.
\end{align}
To properly capture the behavior around this region, the radial integration domain was explicitly split at $k = p$, allowing the global adaptive integration routines to properly sample the integrand without bypassing the regularized singularity.

Finally, rather than recasting the problem as a system of algebraic equations as in~\ref{A}, we solved the coupled integral equations through a damped fixed-point iteration scheme in order to control the oscillatory feedback induced by the vertex dressing. The updated functions at iteration $i+1$ were defined as
\begin{align}
    \Sigma_{i+1}(p) = (1-\omega)\Sigma_i(p) + \omega \Sigma_{calc}(p), \label{B.3} \\[0.3cm]
    A_{i+1}(p) = (1-\omega)A_i(p) + \omega A_{calc}(p), \label{B.4}
\end{align}
where $\Sigma_{calc}(p)$ and $A_{calc}(p)$ denote the outputs of the integral equations evaluated with the functions from the previous iteration. A mixing parameter $\omega=0.3$ was found to provide stable convergence. The procedure was initialized with trial functions, such as $A_0(p)=1.0$ and $\Sigma_0(p)=0.1$, and was iterated until the relative difference between successive steps dropped below a stringent tolerance, yielding a self-consistent solution for the dressed functions.

\section{One-loop correction to the static potential}
\label{C}

For completeness, we briefly derive the corrected static potential associated with the dressed pseudo-Proca propagator of Sec.~\ref{sec:RUnQApprox}. The interaction energy between two static charges is obtained from the time--time component of the propagator in the static limit,
\begin{equation}
V(r)
=
e^{2}
\int \frac{d^{2}k}{(2\pi)^{2}}
\,\Delta_{00}(k_{0}=0,\mathbf{k})\,e^{i\mathbf{k}\cdot\mathbf{r}}.
\label{eq:V_static_def}
\end{equation}

Using the unquenched propagator, Eq.~\eqref{eq:Delta_unquenched}, one finds
\begin{equation}
\Delta_{00}(k_{0}=0,\mathbf{k})
=
\frac{1}{2\sqrt{k^{2}+m^{2}}+\frac{g}{8}\,k},
\label{eq:Delta00_static}
\end{equation}
so that, after angular integration in polar coordinates,
\begin{equation}
V(r)
=
e^{2}
\int_{0}^{\infty}
\frac{k\,dk}{2\pi}
\,
\frac{J_{0}(kr)}
{2\sqrt{k^{2}+m^{2}}+\frac{g}{8}\,k},
\label{eq:V_static_bessel}
\end{equation}
where $J_{0}$ is the zeroth-order Bessel function of the first kind.

\begin{figure}[htpb] 
    \centering
    \includegraphics[width=0.9\linewidth]{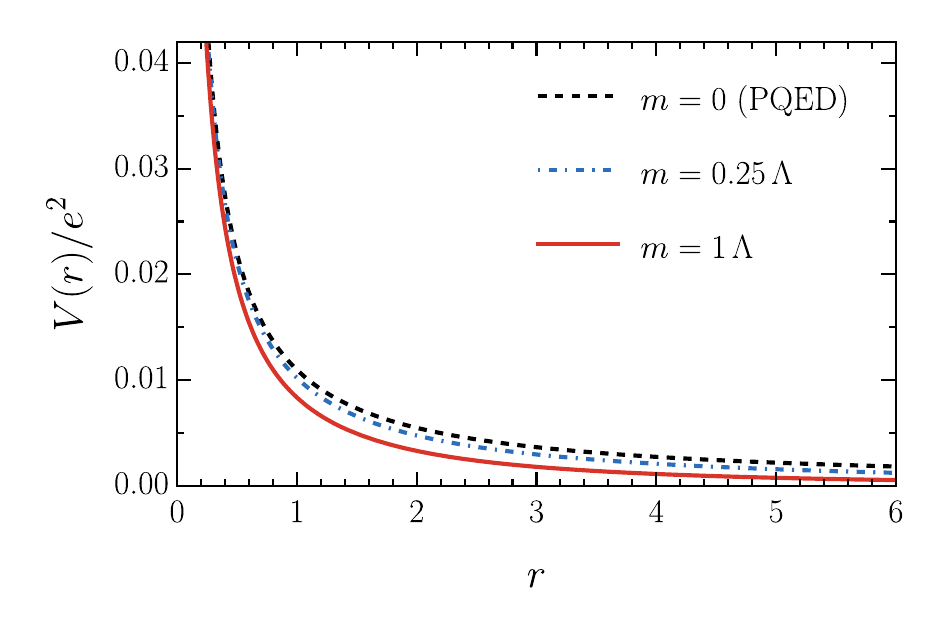}
    \caption{Numerical solution for the corrected static potential in Eq.~\eqref{eq:V_static_bessel} with $g=100$ and different screening scale values: $m=0$ (black), $m=0.25 \, \Lambda$ (blue), and $m=1.0 \, \Lambda$ (red).}
    \label{fig:static_potential}
\end{figure}

From Eq.~\eqref{eq:V_static_bessel}, we can show that the corrected interaction remains screened, with the pseudo-Proca scale $m$ reducing the interaction range. In the limit $m\to 0$, the potential recovers the corrected Coulomb-like behavior of PQED, whereas finite $m$ drives the system toward a Yukawa-like regime. This behavior is illustrated numerically in Fig.~\ref{fig:static_potential}.

A simple interpolating expression that captures the correct asymptotic behavior is
\begin{equation}
V(r)
\approx
\frac{e^{2}}{4\pi r\left(1+\frac{g}{16}\right)}
\exp\left[
-\frac{mr}{1+\frac{g}{16}}
\right].
\label{eq:V_static_ansatz}
\end{equation}
This ansatz reproduces the screened potential in the limit $g\to 0$ and the corrected long-range behavior when $m\to 0$.

Therefore, the one-loop corrected static potential provides a coordinate-space counterpart of the main mechanism discussed in the text: the pseudo-Proca parameter acts as a continuous screening scale that interpolates between long-range and Yukawa-screened interactions. The inclusion of a non-zero fermion mass $M$ modifies the detailed form of the polarization function, but leads to qualitatively similar behavior~\cite{ozela2023effective}.

\bibliographystyle{spphys}      
\bibliography{pseudo_Proca_DMG}   

\end{document}